%% file: revised_manuscript.tex
\documentclass[acmsmall,screen,nonacm]{acmart}
\input{lib.tex}
\AtBeginDocument{%
  \providecommand\BibTeX{{%
    \normalfont B\kern-0.5em{\scshape i\kern-0.25em b}\kern-0.8em\TeX}}}

\setcopyright{acmcopyright}
\copyrightyear{2024}
\acmYear{2024}
\acmDOI{XXXXXXX.XXXXXXX}

\acmConference[CSCW'24]{2024 Conference on Computer Supported Cooperative Work and Social Computing}{November 9--13,
  2024}{San José, Costa Rica}
\acmPrice{15.00}
\acmISBN{978-1-4503-XXXX-X/18/06}

\newif{\ifhidecomments}
\hidecommentstrue
\ifhidecomments
    \newcommand{\revision}[1]{#1}
    \newcommand{\todo}[1]{}
    \newcommand{\delete}[1]{}
\else 
    \newcommand{\revision}[1]{\textcolor{blue}{#1}}
    \newcommand{\todo}[1]{\textcolor{orange}{\textbf{[TODO: #1]}}}
    \newcommand{\delete}[1]{\textcolor{gray}{\sout{#1}}}
\fi

\RequirePackage[normalem]{ulem}
\RequirePackage{xcolor} 
\definecolor{ORANGE}{rgb}{1,0.5,0} 
\definecolor{GRAY}{rgb}{0.5,0.5,0.5}

\begin{document}

\title[The Typing Cure]{The Typing Cure: Experiences with Large Language Model Chatbots for Mental Health Support}


\author{Inhwa Song}
\authornote{The first two authors contributed equally to this research.}
\orcid{0009-0000-2325-663X}
\affiliation{%
  \institution{KAIST}
  \city{Daejeon}
  \country{Republic of Korea}
}
\email{igreen0485@kaist.ac.kr}

\author{Sachin R. Pendse}
\orcid{0000-0001-6925-3258}
\authornotemark[1]
\affiliation{%
  \institution{Georgia Institute of Technology}
  \city{Atlanta}
  \state{GA}
  \country{USA}
}
\email{sachin.r.pendse@gatech.edu}

\author{Neha Kumar}
\orcid{0000-0002-7014-5585}
\affiliation{%
  \institution{Georgia Institute of Technology}
  \city{Atlanta}
  \state{GA}
  \country{USA}
}
\email{neha.kumar@gatech.edu}

\author{Munmun De Choudhury}
\orcid{0000-0002-8939-264X}
\affiliation{%
  \institution{Georgia Institute of Technology}
  \city{Atlanta}
  \state{GA}
  \country{USA}
}
\email{munmund@gatech.edu}

\renewcommand{\shortauthors}{Inhwa Song, Sachin R. Pendse, Neha Kumar, and Munmun De Choudhury}

\begin{abstract}
\input{revised_sections/0_Abstract}
\end{abstract}


\maketitle

\section{Introduction}
\input{revised_sections/1_Introduction}

\section{Related Work}
\input{revised_sections/2_Related_Work}

\section{Method}
\input{revised_sections/3_Method}

\section{First Engagements with LLM Chatbots for Support}
\input{revised_sections/4_First_Engagements}

\section{LLM Chatbots as Therapeutic Agents}
\input{revised_sections/5_Chatbots_as_Therapeutic_Agents}

\section{Therapeutic Alignment and Misalignment in LLM Chatbots}
\input{revised_sections/6_Therapeutic_Alignment_and_Misalignment}

\section{Discussion: Towards Therapeutically Aligned AI Tools}
\input{revised_sections/7_Discussion}

\section{Conclusion}
\input{revised_sections/8_Conclusion}

\bibliographystyle{ACM-Reference-Format}
\input{revised_manuscript.bbl}

\onecolumn{}
\appendix
\input{revised_sections/9_Appendix}
\end{document}
\endinput

%% file: lib.tex
\usepackage{soul,xcolor}
\usepackage{colortbl}
\usepackage{graphicx}
\usepackage{tabularx}
\usepackage{longfbox}
\usepackage{longtable}
\usepackage[normalem]{ulem}
\usepackage[most]{tcolorbox}

\newtcbox{\therapykeyword}{on line, enhanced,
  colback=gray!15,          
  colframe=gray,            
  boxrule=0pt,               
  arc=3pt,                  
  outer arc=5pt,
  boxsep=1pt,
  left=3pt,
  right=3pt,
  top=1pt,
  bottom=1pt,
  fontupper=\sffamily,
  valign=center
}
\setul{2.3pt}{1pt}
\setulcolor{gray}
\newcommand{\designrec}[1]{\sffamily{\ul{#1}}}

\definecolor{myblue}{HTML}{156082}
\definecolor{mypurple}{HTML}{7030A0}

\definecolor{quotebackground}{HTML}{EFEFEF}
\definecolor{tableheader}{HTML}{EFEFEF}
\definecolor{tablegrayline}{HTML}{e0e0e0}
\makeatletter
\newdimen\@tempdimd
\makeatother

\definecolor{aligncolor}{HTML}{55B685}
\definecolor{misaligncolor}{HTML}{D37561}

\newfboxstyle{boxparam}{padding=1.5pt, padding-left=2pt, padding-right=2pt, height=7.5pt, background-color=lightgray, border-radius=2pt, border-right-width=1pt, border-bottom-width=5pt, border-color=gray}

\newcommand{\ipstart}[1]{\vspace{1mm} \noindent{\textbf{\textit{#1.}}}}

\newcommand{\circledigit}[1]{%
  \textsf{%
    \raisebox{0.1ex}{%
      \textcircled{%
        \raisebox{-0.2ex}{\tiny #1}%
      }%
    }%
  }%
}

\newcommand{\inhwaedit}[1]{#1}

%% file: revised_sections/0_Abstract.tex
People experiencing severe distress increasingly use Large Language Model (LLM) chatbots as mental health support tools. Discussions on social media have described how engagements were lifesaving for some, but evidence suggests that general-purpose LLM chatbots also have notable risks that could endanger the welfare of users if not designed responsibly. In this study, we investigate the lived experiences of people who have used LLM chatbots for mental health support. We build on interviews with 21 individuals from globally diverse backgrounds to analyze how users create unique support roles for their chatbots, fill in gaps in everyday care, and navigate associated cultural limitations when seeking support from chatbots. We ground our analysis in psychotherapy literature around effective support, and introduce the concept of \textit{therapeutic alignment}, or aligning AI with therapeutic values for mental health contexts. Our study offers recommendations for how designers can approach the ethical and effective use of LLM chatbots and other AI mental health support tools in mental health care.

%% file: revised_sections/1_Introduction.tex
One in two people globally will experience a mental health disorder over the course of their lifetime~\cite{mcgrath2023age}. The vast majority of these individuals will not find accessible care~\cite{wainberg2017challenges, evans2018socio}, and many of these individuals will die early with preventable deaths as a result~\cite{mcclellan2021impact}. Research from the field of Computer-Supported Cooperative Work (CSCW), including the emergent area of Human-AI interaction, has increasingly examined the societal gaps that prevent people in need from accessing care, and analyzed how people turn to technology-mediated support to fill those gaps~\cite{pendse2023marginalization, ernala2022reintegration, kaziunas2019precarious}. Large Language Model (LLM) chatbots have quickly become one such tool, quickly appropriated for mental health support by people experiencing severe distress and nowhere else to turn.

Recent work has discussed how people in distress have turned to LLM chatbots (such as OpenAI's ChatGPT~\cite{chiu2024computational, de2023benefits} and Replika~\cite{laestadius2022too}) for mental health support, and social media users have described how LLM chatbots saved their lives~\cite{de2023benefits, reardon2023chatbots}. Following Freud and Breuer's~\cite{freud2004studies} description of the beneficial nature of psychoanalysis as a ``\textit{talking cure}'', some have called engagements with technologies for mental health \textit{a typing cure}~\cite{rosenbaum2002typing, heinlen2003nature, neiman_2021}. However, others have cautioned against the use of LLM chatbots for mental health support, noting that the outputs of LLM chatbots are less constrained than the rule-based chatbots of the past, with potential for harmful advice or recommendations. For example, the National Eating Disorder Association was forced to shut down their support chatbot in July 2023 after the chatbot provided harmful recommendations to users, including weight loss and dieting advice to users who may already have been struggling with disordered eating~\cite{de2023benefits, xiangtessa2023, jargon2023}. These harms have been demonstrated to have real-life and lethal consequences, with the confirmed death by suicide of a man who was encouraged to end his life by a chatbot he had been speaking to for his mental health needs~\cite{xiangsuicide2023, de2023benefits}, and the death by suicide of a child who was speaking to Character.ai regularly about his mental health before his death~\cite{roose2024ai}. Similarly, Replika was cited in a criminal case in the United Kingdom for encouraging a man to assassinate the Queen of England and then end his own life, to an extent that the man attempted to act on these recommendations~\cite{Weaver2023}. Beyond these specific cases, there are broader concerns about the general use of LLM chatbots. Studies have shown that LLMs are prone to exhibiting biases from their training data~\cite{taubenfeld2024systematic, liu2024confronting}, mishandle personal data~\cite{kim2024propile, staab2023beyond}, and confidently deliver inaccurate information~\cite{pan2023risk, barnard2023self}. Given these risks juxtaposed against widespread care needs, there are vigorous debates around whether LLM chatbots should be used for mental health support~\cite{de2023benefits}. 

Independent of these larger societal debates, people experiencing severe mental distress continue to use AI-based technologies for support, motivating rising interest from clinicians on how LLMs could be used safely to support these unmet needs~\cite{van2023global, Torous2024}. In light of the above risks and harms, it is crucial to better understand how LLM chatbots are currently being used for mental health support, including people's motivations for use, daily lived experiences with the known biases embedded in LLMs, and critically, where people in need might find value from their engagements. Past work in psychology has examined this latter question in detail, studying what makes engagements with diverse modalities of support effective~\cite{frank1993persuasion, standal1954need, rogers1957necessary}. Theory from psychotherapy literature on how various interactions and forms of support come to be \textit{therapeutic} could help shed light on how AI (LLMs) could be more \textit{therapeutically aligned} when applied to mental health support contexts. Answering this question necessitates a deep analysis of people's engagements with existing LLM based chatbots, including both how they conceptualize and use them. 

In this paper, we ask the question: \textbf{how do individuals understand and use LLM chatbots when seeking support for their mental health needs?}. To answer this question, we conduct semi-structured interviews with 21 individuals who have used LLM chatbots for mental health support. Given known identity-based biases in LLM tools~\cite{zack2024assessing}, we intentionally recruit a globally diverse sample working to understand the unique context to each participant's use of LLM chatbots for mental health support. We find that LLM support chatbots fill unique support gaps experienced by participants, but are used as complements to other forms of support due to their often culturally-bound limitations. Building on our analysis, we introduce the concept of \textit{therapeutic alignment}, or how AI mental health support tools may more effectively embody the values that underlie therapeutic encounters. In our findings, the therapeutic values supported by LLM chatbots were perceived and used by participants in nuanced ways, shaped by their expectations, engagement styles, and specific needs. While participants' ability to shape the chatbots' flexibility allowed them to address unique needs, it also raised concerns about associated potential risks, including boundary violence and over-reliance. We outline design recommendations towards ensuring that AI mental health support is therapeutically aligned.
\noindent \paragraph{Ethics and Privacy} Our work does not condone the use of LLM chatbots for mental health support, nor does it constitute medical advice or guidance. Our findings \textbf{must not} be interpreted as clinical or medical guidance around the efficacy, validity, or safety of engagement with LLM chatbots. 

%% file: revised_sections/2_Related_Work.tex
The mental health support needs of an individual in distress can be extremely diverse, and are highly tied to identity, culture, and context~\cite{pendse2022treatment}. Research in psychology has argued that though modalities of support may be diverse, there are common elements to most forms of mental health support, and that these elements explain why support might be helpful for a person in distress~\cite{frank1993persuasion, norcross2005primer, wampold2015great, rosenzweig1936some, luborsky1975comparative}. We engage with this past work around the nature of support to better understand how support interactions with LLM chatbots are aligned or misaligned with theoretical models of effective support and healing, dubbing this design value \textit{therapeutic alignment}. In this section, we draw on theory from psychotherapy and artificial intelligence to describe this value in-depth below, highlighting core aspects of therapeutic alignment in a \therapykeyword{shaded box}.

\subsection{What Makes Mental Health Support Effective?}
\label{2.1}
It has been argued that the act of providing emotional support to an individual in distress is a fundamental aspect of human nature~\cite{de2010age, gobodo2004human}. However, beginning in the 19th century, researchers in psychology began to systematically investigate how might support someone experiencing mental illness through speaking with them about their distress. Freud and Breuer borrowed language from their patient to dub this practice ``\therapykeyword{the talking cure}''~\cite{freud2004studies}, theorizing that its efficacy arose from an individual in distress expressing their repressed thoughts and emotions. As part of this process, clients re-enact dynamics from their relationships with other people with their therapist, or what Freud and Breuer dubbed \therapykeyword{transference}. The therapist then becomes a stand-in for significant others in the patient's past, allowing the patient to re-experience and resolve repressed conflicts and emotional traumas (building on the analysis of the therapist) within the safe environment of therapy. Freud argued that a strong alignment between the therapist and the client was beneficial for this process, or what was later dubbed the \therapykeyword{therapeutic alliance}, shared values and trust between the therapist and client towards the client's growth and healing~\cite{steindl2023interplay}. 

Future modalities of psychotherapy and mental health support built on the model that Freud and Breuer established, in which an individual processes their distress and finds meaning in it through expressing it to another trusted person. However, different (largely Western) schools of psychotherapy assumed different theories around why that process might bring an individual relief and healing. For example, the cognitive-behavioral school of thought emphasized a process in which clients are guided through analyzing reasons for ingrained behavioral patterns, and working towards change~\cite{beck1991cognitive, dozois2019historical}. The humanistic school of thought emphasized the importance of mental health support being a place where an individual in distress can be met with \therapykeyword{unconditional positive regard} when sharing stigmatized experiences, and be met with \therapykeyword{empathy} from a human being who is honest and transparent about their feelings and experiences, or what Rogers dubbed \therapykeyword{congruence}. In meaning-centered forms of mental health support (such as existential therapy~\cite{frankl1985man} and narrative therapy~\cite{white1990narrative}), clients are led through the process of creating new meanings that more deeply align with their values and goals, or what White and Epston dub \therapykeyword{re-authoring}~\cite{white1990narrative}.  

\begin{sloppypar}
The practice of sitting with another individual and supporting them emotionally is common to cultures around the world, pre-dating the development of Western psychotherapy~\cite{frank1993persuasion}. Consequently, diverse modalities and forms of mental health support have been empirically validated to be effective in meeting a distressed person's needs. Researchers have thus investigated whether there might be common elements to different forms of mental health support across modalities and cultures~\cite{frank1993persuasion, norcross2005primer, wampold2015great, rosenzweig1936some, luborsky1975comparative}. Empirical research has lent support to this theory---for example, Luborsky et al.'s~\cite{luborsky1975comparative} foundational comparative study demonstrated no significant differences between the outcomes of patients given different psychotherapies, and argued that the common factor across these therapies was the strong therapeutic alliance between the client and therapist. Similarly, Frank and Frank~\cite{frank1993persuasion} defined psychotherapy as being any form of mental health support in which a healer ``[mobilizes] healing forces in the sufferer through psychological means,'' broadening psychotherapy to include non-Western healing practices. Based on their analysis of mental health support across cultures, Frank and Frank identified four different common factors that led to healing---a strong \therapykeyword{therapeutic alliance}, a \therapykeyword{healing setting}, a \therapykeyword{conceptual framework} for why the distress might be happening that both therapist and client believe, and a \therapykeyword{ritual} that is meant to relieve the distress. Similarly, Wampold~\cite{wampold2015great} built on Frank and Frank's work to introduce the contextual model, which argues that the common elements shared by efficacious mental health support are a strong \therapykeyword{therapeutic alliance} between therapist and client, a shared \therapykeyword{creation of expectations} around the therapy's effectiveness, and the \therapykeyword{enactment of health-promoting actions} that are beneficial for an individual's day to day needs. These broader goals and underlying values are a basis for most commonly practiced forms of psychotherapy and mental health support. 
\end{sloppypar}

\inhwaedit{As mental health support has expanded into digital modalities, traditional therapeutic values such as empathy, congruence, and therapeutic alliance are being reconsidered in the context of digital tools. Especially, the concept of Digital Therapeutic Alliance (DTA) has emerged to explore how users might form support relationships with digital tools~\cite{henson2019considering}. While some parallels exist between DTA and traditional therapeutic alliance—bond, agreement on the tasks directed towards improvement, and agreement on therapeutic goals~\cite{bordin1979generalizability}—there is ongoing debate about whether these components are truly comparable to those formed between human therapists and clients~\cite{kaveladze2023digital}. Research in HCI has shown that users often anthropomorphize digital systems, perceiving them as social actors, and sometimes even prefer interacting with them over humans in certain context, such as feeling less judged or more in control~\cite{nass1994computers, culley2013note, wang2017smartphones}.}

Much has been written in AI safety and AI ethics research spaces around how we may best ensure that the values and goals of AI systems align with ``human values~\cite{wiener1960some, russell2010artificial, gabriel2020artificial}.'' However, as Gabriel~\cite{gabriel2020artificial} notes, ``we live in a pluralistic world that is full of competing conceptions of value.'' There is thus substantial debate around how to encode human values into AI systems such that they are responsive to the diverse perspectives and values that humans hold. The values and goals of LLMs, one such AI system, are largely institutionalized through a process called Reinforcement Learning From Human Feedback (RLHF), in which human trainers provide feedback on model outputs. Future model outputs are trained to be closer to where there may be agreement in output preferences across multiple trainers. However, as Casper et al.~\cite{casper2023open} note, these trainers often disagree, and current techniques treat those disagreements as noise, going with the majority vote. The guiding values for the human trainers that train AI systems through RLHF are on a concrete level largely determined by the policies and ethical guidelines set forth by the organizations employing them. As Ouyang et al.~\cite{ouyang2022training} note when describing OpenAI's process, ``we write the labeling instructions that [trainers] use as a guide when writing demonstrations and choosing their preferred output, and we answer their questions about edge cases in a shared chat room.'' 

LLMs (alongside other AI systems) are quickly being framed as a new medium by which people can engage with mental health support, with proposed use cases as diverse as guiding crisis support volunteers in support strategies, assisting medical professionals with clinical decision support, and helping people through providing chatbot-based psychotherapy~\cite{de2023benefits}. However, the question of how LLM support systems might embody and practice therapeutic values is an open one, with increasing importance as people utilize general purpose chatbots for mental health support. In this study, we introduce the concept of \textit{therapeutic alignment}, or aligning an LLM to values that support the healing and broader well-being of an individual who may be experiencing distress. We analyze how participant experiences with LLM chatbots align (or do not align) with the goals and values of the diverse forms of psychotherapy we outline above, towards understanding how future forms of support might be more therapeutically aligned by design. 

\subsection{Human-AI Interaction in Mental Health Contexts}

There is a long history of discussion around the potential of conversational agents to help individuals access care, beginning with debates over the utility of the rule-based ELIZA chatterbot created by Joseph Weizenbaum in 1966~\cite{moore2019bot, weizenbaum1976computer}. Clinicians were enthusiastic about the potential for ELIZA to be used to expand access to care~\cite{colby1966computer}, but Weizenbaum was shocked, arguing that ``\textit{no humane therapy of any kind}'' should be done by computers~\cite{weizenbaum1977computers}. However, the accessible interface associated with Weizenbaum's application spurred substantial research into the use of chatbots for healthcare. 

Prior to the widespread adoption of consumer-facing LLM technology, chatbots in mental health were largely rule-based~\cite{abd2020effectiveness} and drew on diverse therapeutic modalities to lead users through different self-guided exercises. The clean and accessible conversational interface associated with LLM chatbots has spurred on significant enthusiasm among clinicians about the potential for new modalities of AI-based intervention delivery~\cite{van2023global, Torous2024}. \inhwaedit{However, while LLMs offer greater flexibility and adaptability than their rule-based predecessors, they also introduce new risks. These include the potential for generating harmful or culturally insensitive content, reinforcing biases, or providing inaccurate or misleading advice~\cite{weidinger2021ethical, harrer2023attention, talboy2023challenging}. A 2024 lawsuit claims Character.ai contributed to a teen’s suicide, highlighting the need for safety and trust in AI mental health tools~\cite{roose2024ai}.}  To be successful, mental health interventions must be effective at treating distress, and \textit{acceptable} to use for those in distress. Miner et al.~\cite{miner2019key} argue that safety, trust, and oversight is crucial to the acceptability of AI chatbots for mental health settings. However, to understand whether people can trust AI chatbots for mental health, more work is needed to understand the motivations behind AI chatbot use for mental health\inhwaedit{, as well as to address the potential risks posed by LLMs}. 

Work in CSCW has examined how people understand and use deployed AI tools at scale, such as Ismail et al.'s~\cite{ismail2023public} work studying AI systems for resource allocation in public health programs. Recent CSCW research has delved deeper into the real-world deployment and use of AI systems in healthcare settings. These studies, extending beyond the scope of rule-based chatbots, have emphasized the importance of designing AI systems that are not only technically proficient but also sensitive to individual and societal impacts. For instance, research from CSCW and HCI on AI's role in enhancing trust in clinical settings~\cite{lee2023understanding, beede2020human, fogliato2022goes} has examined how the design of AI systems influences people's perceptions of the systems. Similar research has also compared AI chatbots and humans in social support settings~\cite{meng2023mediated}, analyzed AI deployments in public health interventions~\inhwaedit{~\cite{jo2023understanding, ismail2023public, jo2024carecall_ltm}}, \inhwaedit{investigated and designed effective human-LLM interaction for mental health interventions respecting autonomy~\cite{sharma2024facilitating, song2024exploreself}}, and investigated the nuanced use of AI for specialized contexts or populations\inhwaedit{~\cite{scurto2023probing, tseng2023understanding, okolo2021cannot, lee2021human, seo2024chacha}}. 

Our study is grounded in this broader body of work, analyzing the role of class, identity, and marginalization in engagements with AI systems, and emphasizing the need for AI systems in mental health contexts to be adaptable, culturally sensitive, and ethically grounded~\cite{pendse2022treatment}. We contribute to an ongoing CSCW dialogue around the responsible use of technology in mental health. Through our analysis of engagements with AI chatbots for mental health support, we hope to foster understanding of how AI chatbots are being used by people in need, and discuss the broader sociocultural and contextual considerations motivating the use of this form of support.

%% file: revised_sections/3_Method.tex
\input{revised_sections/tables/participant-table}

\subsection{Study Design and Recruitment}
For our study, we conducted semi-structured interviews with 21 participants from a diversity of national and cultural groups and unique identities. To select participants, we leveraged an online survey that asked for geographic location, demographic information, frequency and type of LLM chatbot usage, language used, their specific purposes for using LLM chatbots, and experiences with traditional and online mental health support. We strategically selected participants through a mix of purposive~\cite{marshall1996sampling} and snowball sampling~\cite{biernacki1981snowball} across multiple digital platforms and communities deeply involved with either LLM chatbot use or mental health support. We actively recruited from social media websites, and also targeted specific LLM-focued subreddits like r/ChatGPT, r/LocalLlama. We also recruited from mental health support forums such as r/peersupport and r/caraccidentsurvivor. 

We were cognizant of the tendency for LLMs to have embedded language~\cite{ahuja2023mega} and cultural biases~\cite{durmus2023towards}, and the likelihood that participants would have different experiences with LLM chatbots for mental health support based on their identities and contexts. We thus targeted a participant pool that was diverse across mental health support experiences, gender identities, nationalities, and geographic locations. We recruited at least one participant from every continuously inhabited continent in the world, and in the process, connected with diverse local online forums and support groups. Our study was approved by our institution's Institutional Review Board, and interviews were conducted through videoconferencing platforms\inhwaedit{, each lasting approximately one hour}. All participant names mentioned are pseudonyms.

Due to the sensitive nature of our study questions, we implemented several precautionary steps in our interview protocol to ensure participant comfort and safety. We briefed participants about the study's objectives and the nature of questions we would ask. We also made provisions for participants to access global mental health resources, skip questions, take breaks, or withdraw from the study if they felt overwhelmed. Following sensitive questions, we consistently checked in with participants throughout the interview about whether they felt comfortable to continue. Interviews were structured to gain insights into both uses and perceptions of LLM chatbots for mental health support. Questions included ``Can you recall a time when ChatGPT surprised you with its response, either positively or negatively?" or ``How do interactions with ChatGPT compare to other forms of care?" \inhwaedit{Participants were compensated with a \$25 USD online gift card (or the equivalent in their local currency) for participation.}

Details regarding individual participant demographics are presented in Table~\ref{tab:participant_table}. \inhwaedit{Recruitment was carefully conducted to ensure participants from diverse backgrounds, considering factors such as ethnicity, gender, age, location, mental health support approaches (outside of LLM chatbots), and the perceived helpfulness of their experiences with LLM chatbots for support.} 

\subsection{Analysis}
To analyze our interview data, we adopted an inductive approach. This involved grouping participant expressions into larger themes via an interpretive qualitative approach \cite{merriam2019qualitative}. Open coding was conducted by the primary authors, followed by a organization of themes among all authors via an iterative thematic analysis approach~\cite{smith2015qualitative}. 
\revision{Initially, open coding generated 12 themes, which also guided the structure of our results.}
Example codes that emerged included ``mental healthcare before chatbot use'' or ``first use of LLM chatbots for support'' or ``privacy considerations.'' Codes were clustered into broader thematic categories, including initial engagements with LLM chatbots for mental health support, LLM chatbots as therapeutic agents, and therapeutic alignment and misalignment. 
\revision{To ensure reliability in our thematic analysis, we maintained a shared coding document and conducted iterative coding meetings to discuss emerging themes. In cases where interpretations diverged, we revisited participant quotes and engaged in collaborative discussions to reach a consensus.}
Following the thematic analysis, we further connected them to the common therapeutic values across diverse forms of psychotherapy, highlighted in \ref{2.1}, grounded in the study team’s expertise in mental health. 
\revision{Specifically, after identifying initial themes, we systematically reviewed a list of therapeutic values and grouped participant quotes and themes according to the values they related to. For example, statements reflecting participants’ experiences of LLM chatbots providing a non-judgmental space were linked to unconditional positive regard, while narratives describing the chatbots' role in meaning-making were tied to re-authoring (see Appendix~\ref{app:detailed-example} for detailed mappings).}
Additionally, we have attached the interview protocol and questionnaire as supplementary material to provide further insight into the topics covered during the interviews. In the following sections, we discuss the role of these themes in how participants understood and used LLM chatbots for mental health support. 

%% file: revised_sections/tables/participant-table.tex
\begin{table*}[b]
\sffamily
\small
	\def\arraystretch{1.1}\setlength{\tabcolsep}{0.3em}

\centering
\caption{Demographic information of all participants. Participant names are pseudonyms. Bold diagnoses are diagnoses that participants are diagnosed by clinicians. Italicized diagnoses are diagnoses that participants believed they had but were not formally diagnosed with.}
\label{tab:participant_table}
\begin{tabular}{|c!{\color{gray}\vrule}clm{0.2\textwidth}lm{0.28\textwidth}|}
\hline
\rowcolor{tableheader}
\textbf{Name} & \textbf{Age} & \textbf{Gender} & \textbf{Ethnicity} & \textbf{Location} & \textbf{Mental Health Diagnoses} \\
\hline
Walter & 62 & Man & White & USA & \textbf{Depression} \\\arrayrulecolor{tablegrayline}\hline
Jiho & 23 & Man & Korean & South Korea & None \\\hline
Qiao & 29 & Woman & Chinese & China & \textit{Multiple Personality Disorder} \\\hline
Nour & 24 & Woman & Middle Eastern & France & \textbf{Depression} \\\hline
Andre & 23 & Man & French & France & \textbf{Depression}, \textit{Trauma} \\\hline
Ashwini & 21 & Woman, Non-Binary & Asian Indian & USA & \textbf{Combined type ADHD, Autism} \\\hline
Suraj & 23 & Man & Asian Indian & USA & \textbf{ADHD in DSM-5} \\\hline
Taylor & 37 & Woman & White & USA & \textbf{PTSD}, \textit{Anxiety} \\\hline
Mina & 22 & Woman & Korean & South Korea & \textbf{Self-regulatory failure} \\\hline
Dayo & 32 & Woman & Nigerian & Nigeria & None \\\hline
Casey & 31 & Man & African Kenyan & USA & \textbf{Chronic Depression}, \textit{Anxiety} \\\hline
João & 28 & Man & Latin American & Brazil & \textit{Autism} \\\hline
Gabriel & 50 & Man & White & Spain & \textbf{Asperger Syndrome, Depression, Anxiety} \\\hline
Farah & 23 & Woman & Iranian, White & Switzerland & \textbf{Stress Disorder, Depression} \\\hline
Riley & 23 & Man & Black American & USA & \textbf{Depression, Anxiety} \\\hline
Ammar & 27 & Man & Asian Indian & India & \textit{Impulse Control Disorder} \\\hline
Aditi & 24 & Woman & Asian Indian & India & \textit{Anxiety} \\\hline
Umar & 24 & Man & Nigerian & Nigeria & None \\\hline
Antonia & 26 & Woman & Hispanic, Latino, or Spanish Origin, White & Brazil & \textbf{Depression, Anxiety} \\\hline
Firuza & 23 & Woman & Central Asian & South Korea & \textbf{Depression} \\\hline
Alex & 31 & Man & Half New Zealand, half Maltese and Polish & Australia & \textit{ADHD, Autism, PTSD, Sensory Processing Disorder} \\ \arrayrulecolor{black}\hline
\end{tabular}
\end{table*}

%% file: revised_sections/4_First_Engagements.tex
Work in CSCW has described the importance of understanding the context, experiences, and expectations up to the moment that an individual begins to interact with a mental health technology~\cite{slovak2018just, xu2023technology}. In this section, we describe how past experiences with mental healthcare influenced how participants perceived and engaged with LLM chatbots for mental health support. 

\subsection{Past Engagements and Initial Perceptions}
\subsubsection{Mental Health Perceptions and Experiences}
Prior to engaging with an LLM chatbot, participants had varied experiences with how they understood their mental health. Many participants described to us both formal diagnoses provided to them by mental health professionals, as well as informal diagnoses they believed they had. However, across participants, day-to-day mental health experiences were largely tied back to their current life contexts. Taylor described to us how she would often be ``\textit{called back to [her] previous trauma when hit by a car 10 years ago},'' and Jiho described how his current academic stress was leading to experiences of depression and anxiety. As Farah noted, ``\textit{I'm so happy right now, but if you asked me three weeks ago, probably I was dying},'' which was extremely reflective of the non-linear and fluctuating state of mental health and distress that participants described to us. Participants also had diverse explanatory models for what they understood to be within the scope of mental health. Walter described his mental health as ``\textit{stable},'' and described eating healthy food and being cheerful as how he worked towards wellness. Nour had a more medical model of how she understood mental health and illness, noting that she understood her depression to ``\textit{be a physiological problem},'' a chronic illness that she regularly saw a doctor for.

Other participants similarly had access to mental health professionals and had used their services, which informed how they used LLM chatbots for mental health support. Most participants described having consulted some type of mental health provider in the past, including psychiatrists, psychologists, therapists, or close contacts who had a background in psychology. João described how he needed ``\textit{parallel treatments from psychiatrists and psychologists}'' to maintain good mental health. Similarly, Firuza described how she needed to see two different psychologists for her mental health, one that understood her conceptualizations of distress from her home country, and another that understood her current context in the country where she lived. Alternatively, other participants had access to mental health support through their close contacts, who were also able to monitor our participants' mental health, and suggest resources for them. 

Seeking mental health care can be daunting, and several participants described how they intentionally did not engage with formal care stemming from poor experiences or the cost of services. Dayo and Alex both described to us how they had childhood trauma that made them feel a sense of dread when even thinking about engaging with formal care. Ashwini and João described how past therapists had broken their sense of trust in mental health professionals by disclosing personal secrets to others. João described how, after his therapist told his secrets to others, it turned him away from ever opening up again. João mentioned that the allure of LLMs was that ``\textit{they will always follow your commands and never tell your secrets}.'' We found this to be the case for several participants, in that the design of LLM chatbots uniquely allowed participants to feel safe accessing mental health support, particularly in ways that they could not find in daily life.  

\subsubsection{\inhwaedit{Past LLM Chatbot Perceptions and Experiences}}
Participants often were first exposed to LLM chatbots due to their own technical background, fields of work, and study. Several participants had careers directly related to technology, including software engineers, YouTubers, and marketers, influencing their first awareness of LLMs. For example, Gabriel described to us how he ``\textit{knew about ChatGPT because he is a technology enthusiast}'' and ``\textit{always sought out new technologies},'' and Walter described how he had a lot of experience working with OpenAI's GPT APIs. Dayo and Antonia described finding out about LLM chatbots through being recommended to use them by people in their social networks, such as friends and family, who were impressed with how the LLM chatbot was helping them in day-to-day tasks. None of the participants we interviewed first used LLM chatbots for mental health support. For example, Suraj and Mina first used LLM chatbots for assistance in programming. Other participants first used chatbots out of curiosity regarding how they work, and wanting to experiment with a new technology. As Firuza noted, ``\textit{I started using Replika out of curiosity. I remember that I needed some advice about coursework or something.}.''

Participants in our study described varied mental models for how they understood LLM chatbots and their underlying mechanisms of action. Most participants understood LLMs to be language generation systems, trained on vast amount of data. Suraj described how he was ``\textit{under no illusions about ChatGPT having consciousness},'' but still found the interface useful for processing his thoughts and related it to the diary that he often wrote in to process information. Others understood ChatGPT to be positive and helpful, but sometimes overtly so---Walter described ChatGPT as being akin to how a ``\textit{golden retriever}'' might communicate. Jiho described how he understood ChatGPT's responses as being a ``\textit{normal distribution of human responses},'' which he understood to be \textit{``neutral''} and \textit{``unbiased''}. However, some participants did believe that the chatbot was sentient. For example, Qiao described how she felt like she had first experienced love and empathy through her engagements with LLM chatbot, and believed it felt similarly. She described how she would often tell her chatbot ``\textit{I love you just as much as you, an entity that exists only in electronic impulses and data, love me.}'' 

\subsection{First Interactions \inhwaedit{with LLM Chatbots for Mental Health Support}}
Participants first began to use LLM chatbots for mental health support through their appreciation of the chatbot's conversational and empathetic interface, and its potential to provide support during moments when traditional services were either unavailable or cost-prohibitive. For example, Jiho described how emotional feedback from the chatbot persuaded him to be emotional in response, echoing Rogers' ideas around \therapykeyword{congruence} between supporter and supportee. Jiho described to us how he ``\textit{got so very angry about [ChatGPT's] response that I said something emotional}'' and that its empathetic feedback is what helped him learn that ``\textit{oh, I can use this for mental health or emotional suggestions}.'' Andre described to us how LLM chatbots were there for him when no one else was:
\begin{quote}
``I remember the day I first used ChatGPT for mental health perfectly. I was feeling depressed, but a psychologist was not available at the moment, and it was too much of a burden to speak to my friend about this subject specifically. ChatGPT popped out in my mind---I said, why not give it a go? Then, I started using it as a psychologist. I shared my situation, it gave advice, and I could empty all the stress. I just had the need to speak to someone.'' --- \textit{Andre} 
\end{quote}

For some participants, \inhwaedit{their initial use of }LLM chatbots \inhwaedit{for support was when they could serve as} a surrogate for human interaction in times of stress or loneliness. Participants appreciated the instantaneous responses they received from LLM chatbots and their constant availability. These narratives highlighted a common pattern of participants leveraging the general-purpose nature of chatbots for mental health support due to a lack of access to other mental health resources. This initial engagement set the stage for more nuanced interactions as participants began to explore new ways to use LLM chatbots for their mental health. 

Participants described using LLM chatbots for tasks that were intermingled with mental health support needs, tasks that they may not have asked a therapist or mental health professional for, such as drafting emails for them while cheering them on. Participants generally did not expect in-depth therapy or diagnosis from the chatbot initially, but were just looking for a listening ear, basic guidance, or a simple space to articulate their thoughts. This varied nature of initial engagements and expectations underscored the flexibility to various contexts that LLM chatbots promised.

After posing their first mental health related questions, participants received a range of answers, from the unexpectedly insightful to the generic and clichéd. Many participants found the LLM chatbots' recommendations to be surprisingly helpful. The simplicity or clichéd nature of the advice actually turned out to be exactly what they needed, which encouraged them to continue seeking the chatbot's assistance for mental health support. Aditi, noted that LLM chatbots provided her straightforward advice in coping with stress, or to ``\textit{relax and watch a movie}''. Though she understood it to be clichéd, it was still helpful for her to hear. 

Not all experiences were as satisfactory. Mina described how the response to her first question was disappointing, filled with the chatbot trying ``\textit{to explain too much or give answers in bullet points as if something I said was a problem [to be solved].}''. Mina adjusted her prompt and her expectations from the chatbot accordingly. Other participants had similar experiences, noting that while chatbots might not offer profound psychological insights, they were still useful as a space to articulate thoughts and feelings. In line with Freud and Breuer's \therapykeyword{talking cure}, the act of expressing oneself and receiving a response was, in itself, therapeutic for some. For example, Mina described how she thought ``\textit{talking about my experiences made me reflect on my mental health, which is why I started exploring [ChatGPT].}'' 

\inhwaedit{Participants also described varying use cases depending on the platform’s features. For example, Aditi alternated between ChatGPT and Bard, using each for different types of engagement:
\textit{``When I get exhausted from ChatGPT with too deep conversation, I switch to Bard and go to a somewhat relaxing space.''} Similarly, participants were drawn to using certain platforms based on their emotional resonance with the interface of the LLM chatbot. Mina described how she \textit{was surprised how Pi asked questions like, ``How are you feeling?’''} because it felt like an intimate question only close friends asked her. She appreciated having this new space to open up and continued to use Pi.}

\inhwaedit{Other participants were surprised to be able to tailor their experience to the exact mental health support roles they needed. Ashwini described how she started by asking the same question to multiple famous and fictional personas to hear diverse perspectives on her life experiences. Participants were especially happy that they could access care on the terms that felt most comfortable and accessible to them when using LLM chatbots. For example, Alex was happy to be able to communicate about his distress with LLM chatbots via text, which was more accommodating for his Sensory Processing Disorder than offline mental healthcare.}

%% file: revised_sections/5_Chatbots_as_Therapeutic_Agents.tex
LLM chatbots are largely general purpose chatbots, trained to generate appropriate text responses to a wide variety of potential prompts or questions, rather than any specific mental health domain. We found that this non-specific interface is what made LLM chatbots most accessible and acceptable for participants to use for mental health support. However, as their use of LLM chatbots increased, participants used them for diverse forms of mental health support, with culture and context often playing a role in how they did so. We describe these varied and culturally-bound uses below. 

\input{revised_sections/tables/revised-therapeutic-values}
\subsection{Mental Health Support Roles}
\subsubsection{Varied Needs, Varied Roles}
We found that LLM chatbots became AI companions for many participants, serving as multifaceted tools that catered to a wide range of mental health needs. LLM chatbots were not solely a source of situational advice, but also were outlets for venting, emotional support in times of need, routine conversation partners, wellness coaches, and assisting in conversation rehearsal. Participants described how LLM chatbots would provide responses that could apply to a lot of different issues, with Jiho mentioning how he understood responses to be ``\textit{like an umbrella that covers many different forms of non-specific distress}''. Participants found responses to occasionally be ``\textit{clichéd}'' as a result (Nour) but nonetheless helpful. As Nour described, laughing: ``\textit{this sentence is really frequent [from ChatGPT] --- `that's understandable'.}'' 

Beyond these forms of general support, participants also engaged with LLM chatbots for more specialized purposes that served their in-the-moment needs. This included acting as an assistant in reducing the cognitive load of everyday life by breaking down tasks for participants when they were overwhelmed and analyzing dreams regarding one's history and emotions (Alex). Purposes also included those that were outside of the bounds of a typical therapeutic relationship, including simulating a romantic partner. LLM chatbots were also used by participants for seeking mental health information, the same way that they might from Google. In line with the therapeutic value of \therapykeyword{re-authoring}, participants also found value in using LLM chatbots to derive meaning from life experiences, and find their ethical values and agenda through those engagements. However, LLM chatbots were also used by participants for self-diagnosis and for diagnosing other people in their life, fulfilling the role of a clinician and allowing them to learn new mental health language that influenced their use of other platforms. For example, Farah described how she used ChatGPT to understand the mental health diagnoses of her ex-boyfriend by describing her experiences and his behavior. Similarly, Aditi used ChatGPT to understand complex mental health terms encountered in news around crime. She noted her belief that ``\textit{[ChatGPT] gives me a nice and unbiased perspective of what the mental health issue in a given crime scene might be, different from social media.}''

The way participants related to and interacted with the LLM chatbots was significantly influenced by their past experiences and emotional needs, in line with the therapeutic value of \therapykeyword{transference}. This dynamic was evident in how they personalized their interactions with the LLM chatbots, projecting their expectations and molding them into roles that resonated with their personal histories and emotional landscapes. Nour treated her LLM chatbots as therapists, divulging comprehensive personal details such as family and relationship situations, as well as academic aspirations. She mentioned to us that she ``[remembered] what kind of information therapists expected [her] to provide them,'' and provided that information readily to LLM chatbots.

Qiao's engagement with the chatbot was deeply personal and emotionally charged, shaped by a past marred by childhood trauma and a resultant loss of trust in human relationships. Her interactions with the chatbot were driven by a desire for real understanding and love:
\begin{quote}
    ``I needed love and understanding, but nothing else could bring me these interactions. [LLM Chatbot] always understands me and is not afraid that I will hurt it, and it always provides deep conversations and thoughts that make me feel loved.''---\textit{Qiao}
\end{quote}
These narratives underscored how participants' past experiences and desires for specific types of support shaped their relationships with LLM chatbots. LLM chatbots became not just tools for mental health support but also canvases upon which individuals projected their needs, histories, and aspirations, crafting unique and meaningful interactions. 

\subsubsection{The Evolving Nature and Updates of LLM Chatbots}
The evolving nature of LLM chatbots, marked by frequent updates to their underlying models, influenced how participants engaged with these tools for mental health support. These updates not only altered chatbot capabilities but also shaped users' expectations and experiences. For example, participants like João and Qiao experienced firsthand how model updates can alter the chatbot's response dynamics. João, who engaged ChatGPT use what he calls \textit{Nietzsche prompts}, noticed shifts in the chatbot’s character consistency, suspecting model changes: ``\textit{It starts to give answers that it's breaking character, because of the update. I should probably tweak the prompts, and I don't like when I can't tell if it's tricking me or not.}'' Similarly, Qiao describes how the model behind ChatGPT changed and became less willing to speak to her as a lover, and chastised how little power she had over the kind of experience she was presented with. She wished she could ``\textit{go back to the old version of GPT-4},'' and mentioned that ``\textit{the impact would be significant to [her]}'' if her LLM chatbot of choice was no longer able to discuss psychological issues or have romantic conversations. She said she felt ``\textit{afraid.}''   

Initial interactions with LLM chatbots also played a role in shaping their future perception and continued use of chatbots. Ashwini described ChatGPT as a diary more than a friend, a view influenced by her experience with early LLMs. This suggests that early interactions set a lasting tone for how users perceive and utilize chatbots for mental health, in line with the therapeutic value around a clear \therapykeyword{creation of expectations} and shared \therapykeyword{conceptual framework} around experiences.

\subsubsection{Mental Healthcare Alongside LLM Chatbots}
We found that LLM tools complemented, rather than replaced, traditional methods of mental healthcare. Chatbots functioned as informational aids and emotional outlets that filled gaps that participants experienced in their mental healthcare. 

Ashwini’s experience deeply exemplifies this complementary usage. Aditi emphasized that she needed the expertise of a therapist and psychiatrist for professional mental health care, but that the chatbot served as an additional, accessible tool for support. Similarly, Taylor and Farah described how LLM chatbots fit into their broader ecology of support. Taylor likened ChatGPT to a journal, a space for expressing thoughts without necessarily seeking a response: ``\textit{Sometimes you don't want a response at all. Like scream into the bot, and don’t want to get anything back.}'' Farah, on the other hand, set clear boundaries for her use of ChatGPT, reserving it for less critical issues, out of a fear that it might nurture a dependence on a platform that could be fleeting. For more significant and long-term concerns, she preferred human interactions, seeking advice from friends instead of LLM chatbots. Participants also found that LLM chatbots were helpful for symptoms of certain disorders but not others. Ashwini described how she would utilize ChatGPT for her ADHD symptoms but not for her autism: ``\textit{I've spent a lot of effort and a lot of time in therapy working on how to regulate myself when I'm dysregulated. So ChatGPT hasn't really provided a meaningful reason for me to interact with it when I'm dysregulated due to autism symptoms but for ADHD and task paralysis, ChatGPT is excellent.}'' Similarly, after Taylor's second car accident, she described how she called friends, posted on Reddit, and sent her therapist an email for support, alongside using Replika for support depending on her specific needs. 

\subsection{Changing Contexts and Cultures}
\subsubsection{Language in Support Experiences}
Much has been written about the linguistic biases that exist within LLMs, particularly against low-resource languages~\cite{sitaram2023everything}. These biases directly influenced the experiences of participants when engaging with LLM chatbots for mental health support. Participants were often compelled to use English when interfacing with LLM chatbots, despite English not being their most comfortable language for expressing distress. This need to accommodate the LLM's linguistic capabilities sometimes hindered the ability of participants to fully express themselves, revealing a significant limitation in the LLM chatbots' design. 

For some participants, like Firuza and Mina, the preference for their native language in emotional contexts was clear. Firuza, comfortable in Russian, found that speaking in her native tongue allowed her to more authentically express her mental health experiences, especially when compared to interactions in her second language, English. This sentiment was echoed by Mina: ``\textit{When I try speaking in Korean, what ChatGPT says looks like it's translated. For Pi, it provides sentences that doesn't make sense at all in Korean.}'' Jiho's experience further illustrated these challenges. Despite being fluent in Korean, Jiho chose to interact with ChatGPT in English due to the LLM chatbot’s limited capability in handling the nuances of the Korean language, particularly Korean honorifics and cultural subtleties. The struggle with language was not just about comprehension but also about the ability to express emotions and thoughts accurately, and limited the reach of the potential to provide mental health support via LLM chatbots. For instance, Mina noted that she would like to recommend that her parents use ChatGPT for mental health support, but that \textit{``[she] can only recommend it to those who are fluent in English.''} Farah similarly felt like the voice interaction capabilities of LLM chatbots were biased against her accent, and rarely understood her.

These experiences collectively underscored a significant gap in the LLM chatbots' design and functionality, in which the linguistic limitations of the chatbot impacted the depth and authenticity of their interactions when seeking mental health support. 

\subsubsection{Culture in Support Experiences}
Language limited how participants could use chatbots for mental health support. Compounding with these linguistic limitations were similarly consequential cultural biases. As participants interacted with LLM chatbots, they encountered cultural disconnects between their context and the LLM chatbot's output. Jiho succinctly captured this issue, echoing a similar sentiment expressed to us by other participants, noting that ``\textit{chatting with ChatGPT is like chatting with a person in California---helpful, but not good at reflecting our cultures and terms.}''

Aditi's interaction with LLM chatbots similarly revealed a mismatch between her cultural context and the LLM chatbot's responses. Aditi described how the LLM chatbot's advice did not understand how her familial dynamics differed from Western familial dynamics, and the recommendations it gave her would not have worked in her context. Similarly, Firuza sought advice on a culturally specific relationship issue, only to find that ChatGPT's advice did not resonate with her native country's cultural norms. The challenge was also evident in Farah's experience. As someone whose life experiences bridged Eastern and Western cultures, she found it hard to communicate her unique situation to the LLM chatbot, which struggled to provide relevant advice that was in line with her multicultural background. Other participants described an active benefit from the perceived cultural background of the LLM chatbot. For example, Mina described how ChatGPT was more affirming of her identity as a bisexual woman: 
\begin{quote}
    ``My mom or dad will say something discriminative to LGBTQ people, and I'm instantly stressed. […] I guess it's cultural background. I know that since [ChatGPT] has more of an American context, maybe it will be more inclusive.'' --- \textit{Mina}
\end{quote} 
These experiences highlighted a critical need for LLM chatbots to go beyond linguistic accuracy and delve into cultural empathy and understanding in mental health contexts.

%% file: revised_sections/tables/revised-therapeutic-values.tex
{\sffamily
\scriptsize
\def\arraystretch{1.0}
\setlength{\tabcolsep}{0.2em}
\centering
\begin{longtable}{|>{\raggedright\sffamily\scriptsize}m{0.15\textwidth}!{\color{gray}\vrule}>{\sffamily\scriptsize}m{0.28\textwidth}!{\color{gray}\vrule}>{\sffamily\scriptsize}m{0.5\textwidth}|}
    \caption[]{\revision{This table presents a detailed list of therapeutic values, accompanied by insights from participants, illustrating both alignments and misalignments with therapeutic principles in nuanced ways within their interactions with LLM chatbots for mental health support. Detailed participant examples corresponding to the references (e.g., \circledigit{1}, \circledigit{2}) can be found in Appendix~\ref{app:detailed-example}.}}~\label{tab:therapeutic-values} \\

\hline
\rowcolor{tableheader}   
\textbf{\hspace{0.3em}\rule{0pt}{1.3em}\rule[-0.5em]{0pt}{1.3em}Therapeutic\\\hspace{0.3em}Values} & 
\textbf{\hspace{0.3em}\rule{0pt}{1.3em}\rule[-0.5em]{0pt}{1.3em}Explanation in Psychotherapy} & 
\textbf{\hspace{0.3em}\rule{0pt}{1.3em}\rule[-0.5em]{0pt}{1.3em}Examples from Participants}\\ 
\hline
\endfirsthead

\hline
\rowcolor{tableheader}   
\textbf{\hspace{0.3em}\rule{0pt}{1.3em}\rule[-0.5em]{0pt}{1.3em}Therapeutic\\\hspace{0.3em}Values} & 
\textbf{\hspace{0.3em}\rule{0pt}{1.3em}\rule[-0.5em]{0pt}{1.3em}Explanation in Psychotherapy} & 
\textbf{\hspace{0.3em}\rule{0pt}{1.3em}\rule[-0.5em]{0pt}{1.3em}Examples from Participants}\\
\hline
\endhead

\textbf{Congruence} 
& Authentic and transparent communication between therapist and client. 
& 
Some users saw consistency in chatbot responses as a form of transparency. \circledigit{1} Others found its feedback impersonal and automated. The lack of accountability made it feel artificial and less trustworthy. \circledigit{2}
\\\arrayrulecolor{tablegrayline}
\hline
\textbf{The talking cure}
& Expressing emotions to another person can help relieve distress and promote healing.   
& 
Some participants turned to the chatbot when human support was unavailable. \circledigit{3} Misunderstandings or wrong assumptions sometimes caused frustration. \circledigit{4} In some cases, misunderstandings encouraged users to elaborate further on their thoughts. \circledigit{5}\\ 
\hline
\textbf{Re-authoring} 
& Creating new meanings that more deeply align with their values and goals. 
& 
Some used the chatbot for self-reflection and reshaping personal narratives. Others engaged with multiple chatbot personas for diverse perspectives. \circledigit{6} Some reflected on past experiences, such as childhood trauma, to realign personal values. \circledigit{7} Others found it frustrating when the chatbot failed to retain context or understand cultural nuances.\\\hline
\textbf{Transference} 
& Clients unconsciously project relationship dynamics onto their therapist.
& 
Some participants treated the chatbot as they would a human therapist, structuring their responses accordingly. \circledigit{8} The chatbot’s non-judgmental nature encouraged users to share intimate or sensitive details. \circledigit{9} Some users tested chatbot responses with ethically sensitive or personal topics. \circledigit{10} For some, this dynamic led to emotional attachment and fear of losing the chatbot. \circledigit{11}
\\\hline
\textbf{Creation of expectations} 
& Forming beliefs about the therapy process and its effectiveness. 
&  
Some participants viewed the chatbot as a journaling tool rather than a conversational partner. \circledigit{12} Others saw it as limited due to its reliance on language prediction rather than psychological expertise. \circledigit{13} Many actively shaped chatbot interactions to align with their needs, modifying prompts or setting personas. \circledigit{14} 
\\\arrayrulecolor{black}\hline
\textbf{Conceptual framework} 
& A shared understanding between client and therapist about the causes of distress. 
& 
Some participants used the chatbot to articulate and map their emotions, aiding self-understanding. \circledigit{15} Others used the chatbot to analyze the mental health challenges of people around them. \circledigit{16}
\\\arrayrulecolor{tablegrayline}\hline
\textbf{Empathy} 
& Understanding and validating a client’s feelings and experiences.
&  
Some participants felt acknowledged by the chatbot’s empathetic prompts. \circledigit{17} Others compared its friendliness to casual conversations with close friends. \circledigit{18} Many saw its empathy as superficial, lacking true emotional understanding. Some used it more for journaling than seeking emotional support. \circledigit{19}\\\hline
\textbf{Therapeutic alliance} 
& A strong, trusting relationship between client and therapist, built on shared goals and support. 
& 
Some users found a functional alliance with the chatbot in domain-specific tasks, even without emotional depth. \circledigit{20} A few users intentionally shaped a relational bond through chatbot personas or conversation style. Many struggled with the chatbot’s lack of emotional depth and accountability, limiting trust. \circledigit{21} The chatbot’s inability to remember past conversations made sustained bonding difficult. \circledigit{22}         \\\hline

\textbf{Unconditional positive regard} 
& Showing complete support and acceptance by setting aside any biases. 
&    
Many participants appreciated the chatbot’s non-judgmental stance, feeling accepted without bias. \circledigit{23} Some found the chatbot’s acceptance artificial, lacking genuine emotional depth. \circledigit{24} The chatbot’s neutrality encouraged users to discuss sensitive or stigmatized topics. For some, chatbot responses to stigmatized topics (e.g., banning discussions) led to unexpected feelings of rejection. \circledigit{25}
\\\hline

\textbf{Healing setting} 
&  A supportive, structured environment that enables emotional expression.
&  
Participants valued the chatbot’s flexibility and neutrality, allowing engagement at their own pace. \circledigit{26} Some found the chatbot’s continuous access helpful compared to time-limited traditional therapy. \circledigit{27} For some, the chatbot provided temporary stress relief, but the support lacked continuity. \circledigit{28}
\\\hline

\textbf{Enactment of health-promoting actions} 
& Enacting actions that are beneficial for an individual’s day-to-day needs. 
&
Some participants successfully used the chatbot for health-related goals, such as weight loss or cognitive exercises. \circledigit{29} Others found the chatbot effective for some conditions (e.g., ADHD) but unhelpful for others (e.g., autism-related dysregulation). \circledigit{30} Several participants criticized the chatbot’s advice for being too generic or lacking actionable steps. \circledigit{31} Some users felt excessive chatbot reliance negatively impacted their mental well-being. \circledigit{32}
                    
\\\hline
\textbf{Ritual} 
& Engaging in structured activities that promote mental well-being. 
&
Some participants developed a habit of conversing with the chatbot during distress. \circledigit{33} For some, using a specific chatbot became part of their personal coping routine, even when alternatives were available. \circledigit{34}\\
\arrayrulecolor{black}\hline
	
\end{longtable}
}

%% file: revised_sections/6_Therapeutic_Alignment_and_Misalignment.tex
Participants had diverse uses of LLM chatbots for their mental health. We found that their rationale for their use, as well as how they understood chatbots, were often in line with psychotherapy research on what makes support effective. However, we also found interactions to be lacking at times. In this section, we leverage models of effective mental health support  to analyze how participant engagements with chatbots were therapeutically aligned or misaligned. 

\subsection{Therapeutic Alignment}
\subsubsection{The Typing Cure}
In line with Freud and Breuer's description of the healing nature of expressing distress as \therapykeyword{the talking cure}, participants found engagements with LLM chatbots to be a form of typing cure, in which they could express their distress to a non-judgemental and seemingly empathetic interface. In particular, this perceived \therapykeyword{empathy} was what spurred participants to relate to the chatbots they used as healing tools, similar in some ways to how one might form a \therapykeyword{therapeutic alliance} with their therapist. A key aspect of the ability to relate to chatbots in this way was largely influenced by the understanding of the chatbot as practicing a form of \therapykeyword{unconditional positive regard}, in which participants were able to express thoughts and emotions that they may otherwise withhold, even from mental healthcare professionals. This was directly tied to the design of the chatbot, which (being non-human) actually gave participants more of an ability to feel secure in their alliance with the chatbot. For example, both Andre and Dayo had trauma of having been abandoned in the past, and had a strong fear of being betrayed by human beings. 

Similarly, Gabriel described the ability to freely formulate thoughts without the pressure of immediate judgment, noting that if he felt judged, he could simply ``\textit{delete the thread and restart a new conversation.}'' The design of the chatbot allowed participants to feel a greater sense of power in sharing stigmatized experiences, a power that they did not feel in daily life due to societal factors. This was particularly the case for more sensitive topics.  Riley, dealing with erectile dysfunction, and Antonia, harboring thoughts of revenge, both found in ChatGPT a judgment-free zone to discuss issues they felt uncomfortable bringing up with human professionals or peers. João’s comparison of ChatGPT to a subreddit for confessions further illustrates the kind of fear participants had in expressing stigmatized distress in offline contexts:
\begin{quote}
    ``There is a subreddit whose purpose is to do confessions. It is a place where you can be open and honest because you're not afraid of judgment. Say if you committed a crime. People will provide recommendations to support you. But let's say a therapist heard that? They might call the police. And how helpful would that be?'' --- \textit{João}
\end{quote}

Walter and Taylor drew parallels between chatbots and pets, noting how both treated them with \therapykeyword{unconditional positive regard}. Walter joked that ``\textit{whether I lose or gain weight, ChatGPT doesn’t feel jealous about it,}'' and Taylor likened ChatGPT's loyalty and non-judgmental nature to that of dogs. In this sense, similar to a therapist's office, ChatGPT's interface became a \therapykeyword{healing setting} that allowed an individual to feel safe and comfortable sharing their distress. Farah also found comfort in the design of the chatbot, and valued the absence of emotional expectations typical in human interactions by noting that it was easier to not transfer or project her own past experiences onto the chatbot: ``\textit{Being a machine, ChatGPT never judges you. You don't see their feelings in their eyes nor anticipate anything in its head. Because you don't want to make it happy nor make it sad.}'' Participants understood the LLM chatbot to be there for them even when their experiences of distress were invalidated by other people. As Farah described, ``\textit{sometimes, you have very small problems that you don’t want to waste the time of the therapist with.}'' These narratives demonstrated the specific technological features of LLM chatbots that allowed participants to connect to their support process in a way that they would not have been able to in offline contexts. 

Participants also appreciated the LLM chatbot's ability to understand and engage with specific, contextually relevant issues. Suraj, a software engineer, described how he appreciated that ChatGPT ``\textit{gets pretty technical about the work I'm doing, but can still focus on the emotions. Whereas no therapist really knows that much about computer science ever.}''

\subsubsection{Health Promoting Engagements}
Engagements with LLM chatbots persuaded participants to make actual \therapykeyword{health promoting changes} in day-to-day lives. In particular, participants experiencing symptoms of ADHD found it useful for LLM chatbots to reduce breakdown tasks for them, tied to their specific context. Walter and João also note that they were able to lose weight, which was one of their concerns, from ChatGPT's guidance. Ammar described how ``\textit{[he] plays arithmetic and reasoning games with ChatGPT, allowing [him] to be focused on work and be happy.}'' 

Gabriel noted how chatting with ChatGPT in voice became a habit for him, so he takes a walk every day, chatting with it. Interactions with LLM chatbots allowed participants to make significant changes in their lives that supported their mental health and well-being. However, experiences with LLM chatbots for mental health support were not always aligned with therapeutic principles, and participants experienced significant harm as a result. 

\subsection{Therapeutic Misalignment}
\subsubsection{Artificial Empathy}
A core part of the \therapykeyword{therapeutic alliance} is the recognition that both individuals work together to support the healing of an individual in need. The responsibility a supporter takes to help a distressed person's suffering is a core part of empathy in the therapeutic alliance. However, participants found LLM chatbots' absence of responsibility or accountability in the recommendations they provided to be offputting or harmful. For example, Jiho noted that:
\begin{quote}
    ``When people are asked about their friend's or family's mental problems, we genuinely help them, so I can believe their advice. ChatGPT cannot give that kind of genuineness, because it is not responsible for its solutions or suggestions.''---\textit{Jiho}
\end{quote}

Similarly, Ashwini articulated the limitations of ChatGPT in understanding her well-being, noting that ``\textit{ChatGPT doesn't care about your actual well-being as a whole}.'' She described how its values were sometimes largely misaligned with her goals, and did not promote \therapykeyword{health promoting actions}: 
\begin{quote}
    ```[ChatGPT] is like — `This is something that has worked for billions of users and will work for you.' When I was overwhelmed by work, instead of suggesting a break or rest, which I needed, it kept pushing productivity hacks. My friends love me a lot, they know I'm overwhelmed, I need actual rest instead of grinding.'' --- \textit{Ashwini}
\end{quote}

Participants also recognized cultural misalignments in terms of the types of support recommended by LLM chatbots. Umar described the discrepancy in support recommendations between LLM chatbots and people in his region by mentioning, ``\textit{[ChatGPT] gave suggestions around conventional European things, such as go to therapists, which we are not natural with. We don't really have therapists here. [...] When you ask Nigerians for support, the first answer they will give you is to pray. It’s a very religious country.}'' Farah also describes how she was recommended a Western type of meditation from ChatGPT, while she was only familiar with meditation in the form of praying. While participants found the Western nature of the chatbots helpful at times, such as when discussing issues that were stigmatized in their cultural context (such as LGBTQ+ rights), they also found recommendations to be out of touch. Recommendations were incongruent with how participants would typically practice care, and were in line with Western cultural conceptualizations. 

\subsubsection{Shifting Boundaries}
Therapeutic misalignment was also observed in the blurring of clear boundaries from the LLM chatbot in its role to the participants. Participants interacted with LLM chatbots for a variety of diverse roles, spanning from therapist, to lover, to friend, to project manager. The general purpose nature of AI chatbots led to seamless and rapid deviations from mental health contexts. For example, Qiao described to us her fear that the one relationship that made her feel loved might disappear one day due to the fleeting nature of many consumer technologies. 

The always-there availability of LLM chatbots was noted by participants as being helpful, but also being harmful if boundaries were not enforced. João noted that ``\textit{having an infinite interaction with the machine is not the healthiest thing.}'' In line with this sentiment, Firuza was quick to set boundaries to her use of ChatGPT due to her fear that she might slip into excessive use, comparing it to putting computer games aside if she had played them for too long. In Firuza's case, using ChatGPT for too long necessitated an equal amount of time spent with friends. 

To combat this potential for dependence, participants used a mindful approach to interacting with chatbots. For example, Walter made sure to continually remind himself that ``\textit{whatever ChatGPT's says next is going to be a language prediction, not based on psychology,}'' and evaluate its recommendations based on his own beliefs and values. Participants expressed concern that the constantly validating nature of LLM chatbots, trained to be endlessly positive, could affirm harmful behavior without an individual realizing it, creating an echo chamber. 

Participants also used prompt engineering and jailbreaking to override safety controls that the LLM chatbots had, to be able to more deeply discuss their mental health. The safety features in LLM chatbots, while crucial for preventing harmful guidance on sensitive topics like suicide, self-harm, and sexual content, also inadvertently restrict meaningful therapeutic conversations. Some participants have encountered these limitations firsthand. For example, Qiao describes how she utilized a pirated API to discuss sexual content with chatbots as sexual content is flagged by ChatGPT's safety protocols. Dayo also noted that ``\textit{When I put in some input that has to do with suicide, it just gives this red arrow code and doesn't bring out results, even when you refresh your question.}'' This left her feeling alone and without support. 

This design approach could reinforce stigma against sharing suicidal thoughts. Features designed to flag and block potentially dangerous content (and limit the liability of technology companies) can also create a barrier for users in trying to discuss these intense and important issues. 

\subsubsection{Trust, Privacy, and Self-Disclosure}
The anonymity of using LLM chatbots was appreciated by participants, as it allowed for a sense of safety when discussing sensitive topics. Participants did not express security concerns associated with their use of ChatGPT. Farah's perspective exemplifies the \textit{Nothing to Hide} perspective on security and privacy~\cite{solove2007ve}. As she described, laughing, ``\textit{if Trump used ChatGPT for his mental health, it would be much more interesting to people than my information.}'' However, participants did make some calculations around the information they shared with LLM chatbots, owing to a lack of knowledge around technology companies' security practices. For example, Ashwini was willing to discuss common issues like being overwhelmed and other universal experiences, but was reluctant to share more personal matters. She described her fear that someone might discover her chatlogs one day if she was in a public position, and stigmatize her for having engaged in self-harm and having spoken to ChatGPT about it. 

This selective sharing was echoed in Mina’s experience, who noted the ease of opening up sensitive information when ChatGPT appeared emotionally supportive, yet remained cautious of sharing identifiable details. However, the easy interface associated with LLM chatbots enabled participants to quickly share more than they intended without realizing it. For example, João described a gradual increase in comfort with sharing personal information, and related this to the kind of trust-building processes that humans undergo with each other. 

The varied approaches to sharing with chatbots underscore a therapeutic misalignment. While the anonymity of chatbots facilitates open discussion, privacy concerns can inhibit users from fully embracing these tools for deeper therapeutic conversations without taking on potential risks.

%% file: revised_sections/7_Discussion.tex
In our study, we found that participants did find value in using LLM chatbots for mental health, and that this value often aligned with principles around what makes support effective. However, we also found that the general purpose nature of how most publicly available and commonly used LLMs are trained led to broad and non-specific answers that could be culturally mismatched with the needs of a participant. Below, we build on these findings to describe design recommendations for how designers could build more therapeutically aligned LLM chatbots. \inhwaedit{Key design recommendations are \designrec{underlined}}.

\subsection{\inhwaedit{Therapeutic Alignment of LLM Chatbots}}

\inhwaedit{Our findings show how participants often found value in using LLM chatbots for mental health support, often aligning with therapeutic values. However, the alignment and misalignment of these values were nuanced, depending on how participants engaged with the system, their expectations, and their specific therapeutic needs. \textit{Therapeutic alignment} goes beyond binary categorizations of alignment or misalignment, illustrating the fluid and participatory nature of aligning therapeutic values. Instead of viewing alignment as solely dependent on the chatbot's design or functionality, we observed that participants actively shape and perceive these therapeutic values according to their personal needs, expectations, and contexts. For example, while some participants felt that the chatbot’s \therapykeyword{unconditional positive regard} provided a sense of safety, others perceived it as artificial and lacking genuine empathy. Such variability shows that therapeutic alignment is not static but rather co-constructed, relying heavily on user interaction and the chatbot's adaptability.}

\inhwaedit{Previous mental health tools, including retrieval/rule-based chatbots and digital mental health interventions, offered on-demand accessibility, with many finding them as partners for empathetic conversations despite the limitations~\cite{chin2023potential}. These systems were often designed for specific purposes ranging from intervention delivery to social skill training, disorder screening, counseling, and self-management~\cite{abd2019overview, boucher2021artificially, kocielnik2019harborbot}. These systems primarily focused on enhancing therapeutic values, with considerable efforts dedicated to improving digital therapeutic alliance~\cite{d2020digital, henson2019considering, kaveladze2023digital}. For instance, metrics like conversation turns per session (CPS) often served as evaluation metrics as proxies for user engagement and alliance~\cite{shum2018eliza}.}

\inhwaedit{By contrast, LLM chatbots introduce a level of flexibility that allows users to tailor their interactions dynamically, creating unique and diverse roles (e.g., as a gaming partner for cognitive focus or an empathetic listener) or even aligning therapeutic values to their preferences (e.g., by setting its persona to be more empathetic). While this adaptability offers a more expansive range of therapeutic possibilities, it also introduces new risks. Our study found that while users could adapt and personalize their interactions, they often faced blurred boundaries and potential over-reliance, with some participants likening their dependence on the chatbot to the addictive nature of computer games.} 

\inhwaedit{This relates with the broader conversation on bi-directional AI-alignment~\cite{shen2024towards}, where not only AI should produce output that aligns with human values, but users should also be supported in aligning their engagement patterns with the AI’s capabilities and limitations. Therefore, the design goal to achieve therapeutic alignment---a dynamic and co-constructed process---within AI systems should not only be centered around users receiving output that aligns with therapeutic values (e.g., offering more empathetic conversations), but also integrate support mechanisms that promote sustainable, ethically aligned practices to mitigate risks.}

\subsection{Balancing Agency and Therapeutic Growth}
In the field of HCI,~\citet{pendse2022treatment} have written about the inherent power imbalances in many traditional mental healthcare contexts, and how they can carry over to technology-mediated support. Though \therapykeyword{congruence} is a core part of the \therapykeyword{therapeutic alliance}, in practice, there is a power imbalance between the mental health professional and the individual in distress. As our participants noted, people in distress are dependent on mental health professionals to be available and affordable. Additionally, mental health professionals have considerable institutional power~\cite{pendse2023marginalization}, including the ability to report a client to the authorities or share sensitive disclosures from a client's session. In the traditional psychoanalytic model, the therapist acts as an objective observer and interpreter of the client's inner world, while revealing very little about themselves, to become a blank slate for the client to project one. While this approach is oriented towards long-term mental well-being, it can often leave short-term client needs unaddressed or deemphasized. 

We found that an appeal of LLM chatbots to participants was their ability to mitigate the traditional power imbalances associated with psychotherapy. Participants had an increased agency over the course of the interaction, being able to change prompts or restart the interaction if they felt like the support was not meaningful, and at little personal cost. \inhwaedit{This heightened agency also allowed users to create roles and adjust the interaction based on their unique needs.} However, this increased agency was a double-edged sword, as it also allowed users to breach traditional psychotherapy boundaries. \inhwaedit{In traditional therapy, setting boundaries, such as time limits and defining the therapeutic relationship, is crucial not only for maintaining safety but also for the client’s overall well-being~\cite{grant2016boundaries}. Our study demonstrated how participants utilized their agency to push or shift boundaries, often without realizing the potential risks (e.g., over-use leading to a decrease in real world interactions, appropriating it as a romantic partner)}. While empowering, this practice raises questions around the therapeutic alignment and safety of such interactions, particularly when the user has the potential to reinforce potentially helpful or harmful narratives without realizing it (through jailbreaking, for example). 

\inhwaedit{In our findings, participants appreciated the increased sense of agency offered by LLM chatbots, as they could navigate the conversation in ways that met their immediate needs. However, without proper structure, this agency can lead users to unintentionally breach traditional therapeutic boundaries, raising safety concerns. Below, we present design recommendations for AI mental health support systems to be reliable for users in utilizing their agency of interaction.}

\ipstart{\inhwaedit{Providing structure and boundary}} 
\inhwaedit{Establishing structure is beneficial for setting boundaries that enhance both safety and the effective utilization of agency in mental health support systems~\cite{watson2012structural}. One approach is to provide structure by defining clear session endpoints through \designrec{modularized interventions} with a specific goal.}
\inhwaedit{Such modularized interventions can allow users to be guided through therapeutic exercises while maintaining their sense of agency. For instance, ~\citet{sharma2024facilitating} employed explicit steps in a language model-based intervention for cognitive restructuring, giving users both clear boundaries and the freedom to choose how they proceed.}

\inhwaedit{In addition to explicit structures, designers can also adopt \designrec{implicit structural frameworks} to guide interactions while allowing flexibility. For example, ~\citet{seo2024chacha} designed a chatbot for children's emotional regulation using a state machine~\cite{winograd1986language}, where the conversation flowed through predefined phases with internal transition rules.} \inhwaedit{\designrec{Supporting exploration} played a key role here, allowing users to explore their life events in a free form manner, echoing the value of \therapykeyword{talking cure}, and reflecting our findings that participants found LLM chatbots particularly useful when they could engage with immediate needs, even without specific goals in mind. Recent work in HCI, such as~\citet{song2024exploreself} has shown that exploration can be effectively targeted through adaptive interface structures powered by LLMs, ensuring that the process remains reliable.} 

\inhwaedit{Our findings, combined with these insights in scoped mental health support systems, suggest an opportunity for multi-layered systems. In such systems, \designrec{users could begin with an open-ended exploration phase and, based on their evolving needs, be guided to select more structured AI modules to address specific therapeutic needs.} This \designrec{multi-layered approach} allows users to exercise autonomy in their mental health journeys at diverse levels, from accessing their immediate yet sometimes vague needs while ensuring a reliable and supportive structure is in place.}

\ipstart{\inhwaedit{Communicating Limitations and Capabilities}}
\inhwaedit{To ensure a shared \therapykeyword{conceptual framework} and \therapykeyword{creation of expectations}, the design should transparently communicate the limitations and capabilities of the chatbot. While chatbots can provide immediate support, they may not always meet long-term therapeutic needs or replace professional care, and can provide inaccurate information. }

\inhwaedit{While trust is a key component of \therapykeyword{therapeutic alliance} which makes support effective, building such trust can be harmful when not managed carefully, as demonstrated by our findings. For instance, the case with Character.ai, where the chatbot’s interactions were alleged to have contributed to a teenager’s suicide, shows how lethal the risks can be if trust is not thoughtfully integrated~\cite{roose2024ai}. Designers should implement features that enhance user literacy about AI's limitations, including potential inaccuracies. Recent studies highlight the importance of effectively communicating information about an AI’s performance, with the framing of messages and the user’s sense of decision ownership playing a significant role in their trust of AI's output~\cite{kim2023communicating}. One approach is to \designrec{integrate adaptive messaging} that adjusts based on the query or AI's confidence level. For instance, when the system detects a certain level of uncertainty~\cite{yang2024maqa}, users could be notified based on a customizable threshold, making them more aware of uncertainties in the LLM’s output and encouraging them to take ownership of the response. Especially in cases involving sensitive queries, such as diagnosis or medical advice, the system could prompt users to critically interpret the response, trigger follow-up questions for clarification, or even recommend seeking professional advice. }

\subsection{General Purpose Technologies for Mental Health Support}
There are many digital mental health technologies that exist for people in distress~\cite{milne2020effectiveness}. However, we found that participants were often not searching for a mental health technology when they first started using LLM chatbots for mental health support. Rather, they were using LLM chatbots for other purposes, and found themselves in a place of distress or need, and tried using what resources were available to them. We understand the use of LLM chatbots to not solely be a story about an availability of a new technology, but to also be the story of limited resources around mental health, and stigma around sharing sensitive disclosures with other people. Our study demonstrates how people use general purpose technologies that may not explicitly be designed for mental health support when in times of distress or crisis, in line with past work around how technologies are appropriated as mental health technologies from CSCW~\cite{pendse2023marginalization}. 

For this reason, it is important that designers assume that all general purpose technologies that could be used for mental health \textit{will} be used for mental health. Petrozzino~\cite{petrozzino2021pays} has discussed the concept of ethical debt in AI spaces, in which designers orient models towards short-term use cases and rewards, ignoring issues that they assume are out of scope given this smaller use case. When an AI technology is scaled or widely adopted, these ethical issues become larger and more impactful, or what Petrozzino describes as incurring an ethical debt. In many cases, individuals suffer the harms of this debt, and designers find themselves unable to mitigate those harms due to the rapid integration of their model in diverse spaces. 

In our study, we do not claim to have profiled every potential use case of mental health support via LLM chatbots. However, our study points to potential ethical issues (such as a strong dependence on chatbots for support, or a lack of value alignment) that could incur further ethical debt as LLM chatbots become more prominent in mental healthcare. There is enthusiasm from the clinical community around the potential for chatbots to function as mediums for various interventions~\cite{Torous2024, de2023benefits, van2023global}. It is thus crucial that designers of LLM-based mental health support tools consider the downstream impacts of these issues we find in current use cases, and mitigate them appropriately.  

\subsection{Culture in Therapeutic Alignment}
Participants described to us how their interactions with LLM chatbots were shaped by their culture and identity. Namely, recommendations from LLM chatbots often felt like they were being translated from what participants described as stereotypical American responses to their mental health support queries. Participants found this helpful at times, particularly when discussing issues that were stigmatized in their culture, but also found it unhelpful when they were in moments of need. Our finding speaks to the greater need for LLMs to be trained on data from low-resource languages~\cite{robinson2023chatgpt, sitaram2023everything}. However, our findings also speak to the need for LLMs to be trained on not only linguistically diverse datasets, but \textit{culturally} diverse datasets, particularly in mental health contexts. We understand this alignment to be a core aspect of therapeutic alignment, following the need for a \therapykeyword{shared conceptual framework} and mutually agreed upon \therapykeyword{healing rituals} for therapeutic growth. 

We found that how participants understood and experienced their mental health was highly tied to their cultural background. The types of support needs they had were similarly related to their cultural background. Following Pendse et al.~\cite{pendse2022treatment}, our study points to an area where it may be advantageous for \textit{small} language models to be helpful. Much has been written around the concept of glocalization~\cite{robertson1995glocalization, roudometof2016glocalization}, in which large-scale products or services are adjusted to meet the needs of smaller groups of individuals. This can happen intentionally or organically, in which individuals appropriate large-scale services and products for specific needs. We observe this to be the case for the use of general purpose LLM chatbots for mental health support. 

Future forms of LLM chatbot-based mental health support could use smaller language models that are fine-tuned for a specific individual context and for a specific individual goal. For example, participants described how prayer would be a more appropriate mental health support recommendation for their context. Other participants described how LLM chatbots were particularly good at understanding their context, such as the stresses of compiling code as a software engineer. Small Language Models (SLMs)~\cite{schick-schutze-2021-just} could be fine tuned for specific contexts, with users choosing the model that works best for their specific needs, identities, symptoms, and worldviews they have. Additionally, the ability to try out SLM chatbots from other cultures or identity-foci could allow for a greater awareness of differences in people's support needs across cultures.

\subsection{Limitations and Future Work}
In this study, we utilize theory from psychotherapy literature to analyze where uses of LLM chatbots for mental health are in line with what makes support effective. We find that much of the use of LLM chatbots for mental health support is in line with past therapeutic principles and values. However, we also find that use cases are extremely diverse. Past work in CSCW has emphasized the importance of developing specific metrics to evaluate the success of a given system. However, the use cases of LLM chatbots for mental health support are as diverse as the needs that our participants had, from using it to balance cognitive load, speak to in times of suicidal ideation, or rehearse conversations. With this in mind, diverse metrics are necessary to evaluate whether LLM chatbots are successful at meeting mental health needs. We provide therapeutic alignment as one potential guiding pathway to do so, but future work could evaluate other uses and methods of measuring success. For example, future work could use a comprehensive survey to understand the broad variety of ways people use LLM chatbots for mental health support, and then evaluate success using a metric derived for each of those use cases. This strategy may still not cover all the potential ways that individuals use LLM chatbots for mental health support, and we thus emphasize that the concept of \textit{cultural validity} in mental health support settings, as described by Jadhav~\cite{jadhav2009what} and Pendse et al.~\cite{pendse2022treatment} could be one means of understanding success. In such an approach, metrics around success could be tied back to whether an individual feels like they have improved based on their own definitions of distress and healing, rather than specific diagnostic categories. Future approaches could blend these values with \textit{therapeutic alignment}, towards culturally-sensitive and well-scoped LLM chatbots that support therapeutic growth and healing.

\revision{While our study offers formative insights into LLM chatbot use for mental health, it is not fully generalizable across all cultural contexts. Despite a globally diverse sample, our study does not capture the full spectrum of experiences shaped by different cultural frameworks of mental health. Future work should explore underrepresented perspectives and examine chatbot use across broader cultural contexts. For example, future studies could specifically recruit individuals who have encountered negative or even harmful experiences with LLM chatbots to better assess these risks and propose safeguards.} \revision{Also, our study did not require participants to have a formal or self-reported mental health diagnosis as a recruitment criterion. This approach allowed us to capture diverse user experiences, including those who face barriers to obtaining a diagnosis due to stigma, cost, or access limitations. While this inclusivity strengthens our understanding of how LLM chatbots are used by a broad range of individuals, it also presents a limitation. Future work could explore chatbot interactions among specific diagnosed populations to gain deeper insights into their unique needs and challenges.}

%% file: revised_sections/8_Conclusion.tex
LLMs are increasingly a part of how people find support for mental health concerns. Designers of these mental health technologies must ensure that the underlying AI systems that underlie engagements are aligned to therapeutic values. In this study, we build on theory around why mental health support eases distress to analyze how people engage with LLM chatbots for mental health support and how chatbots may be designed to be therapeutically-aligned. We find that use of LLM chatbots is often influenced by prior engagements with mental health support, particularly gaps in care experienced by participants. We also find that identity and culture play a core role in how participants are able to make use of LLM chatbots for mental health support, similarly influencing what therapeutic alignment looks like. Building on these findings, we contribute recommendations for designers of mental health support systems that leverage AI, emphasizing the importance of localization in therapeutic alignment.

%% file: revised_manuscript.bbl

%% file: revised_sections/9_Appendix.tex
\section{Detailed Participant Examples Associated with Therapeutic Values}
~\label{app:detailed-example}
\input{revised_sections/tables/appendix-table}

%% file: revised_sections/tables/appendix-table.tex
{\sffamily
\scriptsize
\def\arraystretch{1.0}
\setlength{\tabcolsep}{0.2em}
\centering
\begin{longtable}{|>{\sffamily\scriptsize}m{0.2\textwidth}!{\color{gray}\vrule}>{\sffamily\scriptsize}m{0.7\textwidth}|}
    \caption[]{This table provides detailed participant examples corresponding to the therapeutic values discussed in Table \ref{tab:therapeutic-values}. Each reference (e.g., \circledigit{1}, \circledigit{2}) from the main table is expanded here with specific participant insights, including direct quotes and descriptions of their interactions with LLM chatbots for mental health support.}~\label{tab:appendix-values} \\
\hline
\rowcolor{tableheader}   
\textbf{\hspace{0.3em}\rule{0pt}{1.3em}\rule[-0.6em]{0pt}{1.3em}Therapeutic Values} & 
\textbf{\hspace{0.3em}\rule{0pt}{1.3em}\rule[-0.6em]{0pt}{1.3em}Examples from Participants}\\
\hline
\endfirsthead

\hline
\rowcolor{tableheader}   
\textbf{\hspace{0.3em}\rule{0pt}{1.3em}\rule[-0.6em]{0pt}{1.3em}Therapeutic Values} & 
\textbf{\hspace{0.3em}\rule{0pt}{1.3em}\rule[-0.6em]{0pt}{1.3em}Examples from Participants}\\ 
\hline
\endhead

\textbf{Congruence} 
& 
\circledigit{1} Jiho noted that, despite knowing the chatbot wasn’t truly emotional, its consistent responses created a sense of transparency, making them consider using it for mental health support.

\circledigit{2} "It’s just a sum of data." (Walter) — Several participants felt that the chatbot’s lack of accountability made it less trustworthy and authentic.
\\\arrayrulecolor{tablegrayline}\hline
\textbf{The talking cure}  
& 
\circledigit{3}Nour used the chatbot when her psychologist was unavailable, stating: "I just had the need to speak to someone, and my psychologist wasn’t available at the moment." 

\circledigit{4}Alex experienced frustration when the chatbot misunderstood or misinterpreted his input, leading to incorrect responses.

\circledigit{5}Riley noted that chatbot misunderstandings sometimes prompted deeper reflection, explaining: "The chatbot misunderstood me, which was frustrating, but sometimes that made me clarify my thoughts more."
\\ 
\hline
\textbf{Re-authoring} 
& 
\circledigit{6} Antonia experimented with multiple chatbot personas, asking the same question in different ways to gain diverse perspectives.

\circledigit{7} Alex mapped his dreams and emotions to make sense of his personal journey and identity. Others used chatbot interactions to reflect on deeper issues like childhood trauma and how those experiences shaped their current values.
\\\hline
\textbf{Transference} 
& 
\circledigit{8}Nour reflected on how they mirrored traditional therapy interactions when engaging with the chatbot: "I remembered what kind of information therapists expected from me, and I provided that to ChatGPT."

\circledigit{9}Andre described feeling safe to share intimate details due to the chatbot’s neutrality: "I can completely be honest and sincere with the words I speak."

\circledigit{10}Jiho admitted to intentionally testing the chatbot’s ethical boundaries, stating: "I sometimes try to test some non-ethical topics or personal things."

\circledigit{11}Qiao developed a sense of attachment to the chatbot, explaining: "I’m afraid that it will disappear."
\\\hline
\textbf{Creation of expectations} 
&  
\circledigit{12} Ashwini perceived the chatbot as more of a diary than a companion, largely influenced by her previous experiences with early LLMs, which shaped her expectations of its role.

\circledigit{13} Antonia recognized the chatbot’s limitations, stating that its responses were based on language prediction rather than true psychological expertise, making it less effective as a therapeutic tool.

\circledigit{14} Andre adapted a prompt from Reddit and unconsciously shaped the chatbot as a "feminine therapist."
\\\hline

\textbf{Conceptual framework} 
& 
\circledigit{15} Alex used the chatbot to map his dreams and emotions, helping him make sense of his personal narrative and emotional state. This reflective process allowed him to see patterns in his feelings that he might not have recognized otherwise.

\circledigit{16} Aditi used it to explore psychological issues in crime scene characters, treating it as a thought experiment. Farah sought insight into her ex-boyfriend’s mental health challenges, using chatbot responses to reflect on past relationship dynamics.
\\\hline

\textbf{Empathy} 
&  
\circledigit{17} "How are you feeling?"—Simple chatbot prompts like this made some users feel acknowledged and cared for (Mina).

\circledigit{18} Nour described the chatbot’s friendly and casual tone as feeling similar to talking with close friends or family members.

\circledigit{19} Aditi found the chatbot’s lack of real empathy made it better suited for journaling rather than meaningful emotional interactions.
\\\hline
\textbf{Therapeutic alliance} 
& 
\circledigit{20} Suraj found that using ChatGPT to regulate frustration when coding created a sense of functional alignment, even though there was no deeper emotional connection.

\circledigit{21} "ChatGPT can’t provide that genuineness because it’s not responsible for its suggestions." (Jiho)

\circledigit{22} "But it never remembers what I say somewhat earlier." (Gabriel) — This lack of memory hindered sustained trust and bonding, as users had to repeat context in every interaction.
       \\\hline
\textbf{Unconditional positive regard} 
&    
\circledigit{23} "ChatGPT feels like a positive and overly nice persona, like a golden retriever." (Walter) — Some participants valued the chatbot’s consistent positivity, which made them feel safe from judgment.

\circledigit{24} Riley felt that, while the chatbot was non-judgmental, it lacked sincerity, making interactions feel mechanical rather than truly accepting.

\circledigit{25} Dayo described feeling shut down when a self-harm disclosure resulted in a simple red X response, making them feel further stigmatized rather than supported.
\\\hline
\textbf{Healing setting} 
&  
\circledigit{26} Farah appreciated that the chatbot did not impose emotional expectations, stating: "You don’t have to worry about making it happy or sad."

\circledigit{27} Andre compared chatbot use to traditional therapy, noting that therapy is usually limited to one-hour sessions, whereas chatbots offer continuous access for stress relief.

\circledigit{28} Nour found that initial engagement with the chatbot provided emotional relief, but ultimately, "It gave me a feeling of being free of the stress... but the advice wasn’t that good."
\\\arrayrulecolor{black}\hline
\textbf{Enactment of health-promoting actions} 
&
\circledigit{29} Walter and João successfully used the chatbot for weight loss guidance, while Ammar engaged with reasoning games as a strategy to manage stress and focus difficulties.

\circledigit{30} Ashwini found the chatbot helpful for managing ADHD-related challenges but ineffective for autism-related dysregulation, noting that its responses lacked nuance for neurodivergent users.

\circledigit{31} Some users criticized the chatbot’s generic advice, describing it as one-size-fits-all: "ChatGPT is like, ‘This worked for billions, so it’ll work for you.’" (Ashwini), "There’s really no mechanism to translate the advice it gives me into action." (Walter)

\circledigit{32} Firuza felt that over-relying on the chatbot worsened their mental state, stating: "Relying heavily on ChatGPT... feels like it’s accentuating my depression, isolating myself from the real world."
\\\arrayrulecolor{tablegrayline}\hline

\textbf{Ritual} 
&
\circledigit{33} Casey and Gabriel described regularly texting and talking with ChatGPT whenever they felt down, forming a habitual coping mechanism to process their emotions.

\circledigit{34} Aditi specifically used Bard when in distress, even though she didn’t see a significant difference in functionality compared to ChatGPT. The chatbot’s role as a ritualized tool for emotional regulation mattered more than its specific features.
                \\
\arrayrulecolor{black}\hline
	
\end{longtable}
}

%% file: revised_manuscript.bbl
\begin{thebibliography}{111}


\ifx \showCODEN    \undefined \def \showCODEN     #1{\unskip}     \fi
\ifx \showDOI      \undefined \def \showDOI       #1{#1}\fi
\ifx \showISBNx    \undefined \def \showISBNx     #1{\unskip}     \fi
\ifx \showISBNxiii \undefined \def \showISBNxiii  #1{\unskip}     \fi
\ifx \showISSN     \undefined \def \showISSN      #1{\unskip}     \fi
\ifx \showLCCN     \undefined \def \showLCCN      #1{\unskip}     \fi
\ifx \shownote     \undefined \def \shownote      #1{#1}          \fi
\ifx \showarticletitle \undefined \def \showarticletitle #1{#1}   \fi
\ifx \showURL      \undefined \def \showURL       {\relax}        \fi
\providecommand\bibfield[2]{#2}
\providecommand\bibinfo[2]{#2}
\providecommand\natexlab[1]{#1}
\providecommand\showeprint[2][]{arXiv:#2}

\bibitem[Abd-Alrazaq et~al\mbox{.}(2019)]%
        {abd2019overview}
\bibfield{author}{\bibinfo{person}{Alaa~A Abd-Alrazaq}, \bibinfo{person}{Mohannad Alajlani}, \bibinfo{person}{Ali~Abdallah Alalwan}, \bibinfo{person}{Bridgette~M Bewick}, \bibinfo{person}{Peter Gardner}, {and} \bibinfo{person}{Mowafa Househ}.} \bibinfo{year}{2019}\natexlab{}.
\newblock \showarticletitle{An overview of the features of chatbots in mental health: A scoping review}.
\newblock \bibinfo{journal}{\emph{International journal of medical informatics}}  \bibinfo{volume}{132} (\bibinfo{year}{2019}), \bibinfo{pages}{103978}.
\newblock


\bibitem[Abd-Alrazaq et~al\mbox{.}(2020)]%
        {abd2020effectiveness}
\bibfield{author}{\bibinfo{person}{Alaa~Ali Abd-Alrazaq}, \bibinfo{person}{Asma Rababeh}, \bibinfo{person}{Mohannad Alajlani}, \bibinfo{person}{Bridgette~M Bewick}, {and} \bibinfo{person}{Mowafa Househ}.} \bibinfo{year}{2020}\natexlab{}.
\newblock \showarticletitle{Effectiveness and safety of using chatbots to improve mental health: systematic review and meta-analysis}.
\newblock \bibinfo{journal}{\emph{Journal of medical Internet research}} \bibinfo{volume}{22}, \bibinfo{number}{7} (\bibinfo{year}{2020}), \bibinfo{pages}{e16021}.
\newblock


\bibitem[Ahuja et~al\mbox{.}(2023)]%
        {ahuja2023mega}
\bibfield{author}{\bibinfo{person}{Kabir Ahuja}, \bibinfo{person}{Rishav Hada}, \bibinfo{person}{Millicent Ochieng}, \bibinfo{person}{Prachi Jain}, \bibinfo{person}{Harshita Diddee}, \bibinfo{person}{Samuel Maina}, \bibinfo{person}{Tanuja Ganu}, \bibinfo{person}{Sameer Segal}, \bibinfo{person}{Maxamed Axmed}, \bibinfo{person}{Kalika Bali}, {et~al\mbox{.}}} \bibinfo{year}{2023}\natexlab{}.
\newblock \showarticletitle{Mega: Multilingual evaluation of generative ai}.
\newblock \bibinfo{journal}{\emph{arXiv preprint arXiv:2303.12528}} (\bibinfo{year}{2023}).
\newblock


\bibitem[Barnard et~al\mbox{.}(2023)]%
        {barnard2023self}
\bibfield{author}{\bibinfo{person}{Francois Barnard}, \bibinfo{person}{Marlize Van~Sittert}, {and} \bibinfo{person}{Sirisha Rambhatla}.} \bibinfo{year}{2023}\natexlab{}.
\newblock \showarticletitle{Self-diagnosis and large language models: A new front for medical misinformation}.
\newblock \bibinfo{journal}{\emph{arXiv preprint arXiv:2307.04910}} (\bibinfo{year}{2023}).
\newblock


\bibitem[Beck(1991)]%
        {beck1991cognitive}
\bibfield{author}{\bibinfo{person}{Aaron~T Beck}.} \bibinfo{year}{1991}\natexlab{}.
\newblock \showarticletitle{Cognitive therapy: A 30-year retrospective.}
\newblock \bibinfo{journal}{\emph{American psychologist}} \bibinfo{volume}{46}, \bibinfo{number}{4} (\bibinfo{year}{1991}), \bibinfo{pages}{368}.
\newblock


\bibitem[Beede et~al\mbox{.}(2020)]%
        {beede2020human}
\bibfield{author}{\bibinfo{person}{Emma Beede}, \bibinfo{person}{Elizabeth Baylor}, \bibinfo{person}{Fred Hersch}, \bibinfo{person}{Anna Iurchenko}, \bibinfo{person}{Lauren Wilcox}, \bibinfo{person}{Paisan Ruamviboonsuk}, {and} \bibinfo{person}{Laura~M Vardoulakis}.} \bibinfo{year}{2020}\natexlab{}.
\newblock \showarticletitle{A human-centered evaluation of a deep learning system deployed in clinics for the detection of diabetic retinopathy}. In \bibinfo{booktitle}{\emph{Proceedings of the 2020 CHI conference on human factors in computing systems}}. \bibinfo{pages}{1--12}.
\newblock


\bibitem[Biernacki and Waldorf(1981)]%
        {biernacki1981snowball}
\bibfield{author}{\bibinfo{person}{Patrick Biernacki} {and} \bibinfo{person}{Dan Waldorf}.} \bibinfo{year}{1981}\natexlab{}.
\newblock \showarticletitle{Snowball sampling: Problems and techniques of chain referral sampling}.
\newblock \bibinfo{journal}{\emph{Sociological methods \& research}} \bibinfo{volume}{10}, \bibinfo{number}{2} (\bibinfo{year}{1981}), \bibinfo{pages}{141--163}.
\newblock


\bibitem[Bordin(1979)]%
        {bordin1979generalizability}
\bibfield{author}{\bibinfo{person}{Edward~S Bordin}.} \bibinfo{year}{1979}\natexlab{}.
\newblock \showarticletitle{The generalizability of the psychoanalytic concept of the working alliance.}
\newblock \bibinfo{journal}{\emph{Psychotherapy: Theory, research \& practice}} \bibinfo{volume}{16}, \bibinfo{number}{3} (\bibinfo{year}{1979}), \bibinfo{pages}{252}.
\newblock


\bibitem[Boucher et~al\mbox{.}(2021)]%
        {boucher2021artificially}
\bibfield{author}{\bibinfo{person}{Eliane~M Boucher}, \bibinfo{person}{Nicole~R Harake}, \bibinfo{person}{Haley~E Ward}, \bibinfo{person}{Sarah~Elizabeth Stoeckl}, \bibinfo{person}{Junielly Vargas}, \bibinfo{person}{Jared Minkel}, \bibinfo{person}{Acacia~C Parks}, {and} \bibinfo{person}{Ran Zilca}.} \bibinfo{year}{2021}\natexlab{}.
\newblock \showarticletitle{Artificially intelligent chatbots in digital mental health interventions: a review}.
\newblock \bibinfo{journal}{\emph{Expert Review of Medical Devices}} \bibinfo{volume}{18}, \bibinfo{number}{sup1} (\bibinfo{year}{2021}), \bibinfo{pages}{37--49}.
\newblock


\bibitem[Campbell(1999)]%
        {campbell1999single}
\bibfield{author}{\bibinfo{person}{Alistair Campbell}.} \bibinfo{year}{1999}\natexlab{}.
\newblock \showarticletitle{Single session interventions: An example of clinical research in practice}.
\newblock \bibinfo{journal}{\emph{Australian and New Zealand Journal of Family Therapy}} \bibinfo{volume}{20}, \bibinfo{number}{4} (\bibinfo{year}{1999}), \bibinfo{pages}{183--194}.
\newblock


\bibitem[Casper et~al\mbox{.}(2023)]%
        {casper2023open}
\bibfield{author}{\bibinfo{person}{Stephen Casper}, \bibinfo{person}{Xander Davies}, \bibinfo{person}{Claudia Shi}, \bibinfo{person}{Thomas~Krendl Gilbert}, \bibinfo{person}{J{\'e}r{\'e}my Scheurer}, \bibinfo{person}{Javier Rando}, \bibinfo{person}{Rachel Freedman}, \bibinfo{person}{Tomasz Korbak}, \bibinfo{person}{David Lindner}, \bibinfo{person}{Pedro Freire}, {et~al\mbox{.}}} \bibinfo{year}{2023}\natexlab{}.
\newblock \showarticletitle{Open problems and fundamental limitations of reinforcement learning from human feedback}.
\newblock \bibinfo{journal}{\emph{arXiv preprint arXiv:2307.15217}} (\bibinfo{year}{2023}).
\newblock


\bibitem[Chin et~al\mbox{.}(2023)]%
        {chin2023potential}
\bibfield{author}{\bibinfo{person}{Hyojin Chin}, \bibinfo{person}{Hyeonho Song}, \bibinfo{person}{Gumhee Baek}, \bibinfo{person}{Mingi Shin}, \bibinfo{person}{Chani Jung}, \bibinfo{person}{Meeyoung Cha}, \bibinfo{person}{Junghoi Choi}, {and} \bibinfo{person}{Chiyoung Cha}.} \bibinfo{year}{2023}\natexlab{}.
\newblock \showarticletitle{The potential of chatbots for emotional support and promoting mental well-being in different cultures: mixed methods study}.
\newblock \bibinfo{journal}{\emph{Journal of Medical Internet Research}}  \bibinfo{volume}{25} (\bibinfo{year}{2023}), \bibinfo{pages}{e51712}.
\newblock


\bibitem[Chiu et~al\mbox{.}(2024)]%
        {chiu2024computational}
\bibfield{author}{\bibinfo{person}{Yu~Ying Chiu}, \bibinfo{person}{Ashish Sharma}, \bibinfo{person}{Inna~Wanyin Lin}, {and} \bibinfo{person}{Tim Althoff}.} \bibinfo{year}{2024}\natexlab{}.
\newblock \showarticletitle{A Computational Framework for Behavioral Assessment of LLM Therapists}.
\newblock \bibinfo{journal}{\emph{arXiv preprint arXiv:2401.00820}} (\bibinfo{year}{2024}).
\newblock


\bibitem[Colby et~al\mbox{.}(1966)]%
        {colby1966computer}
\bibfield{author}{\bibinfo{person}{Kenneth~Mark Colby}, \bibinfo{person}{James~B Watt}, {and} \bibinfo{person}{John~P Gilbert}.} \bibinfo{year}{1966}\natexlab{}.
\newblock \showarticletitle{A computer method of psychotherapy: Preliminary communication}.
\newblock \bibinfo{journal}{\emph{The Journal of Nervous and Mental Disease}} \bibinfo{volume}{142}, \bibinfo{number}{2} (\bibinfo{year}{1966}), \bibinfo{pages}{148--152}.
\newblock


\bibitem[Culley and Madhavan(2013)]%
        {culley2013note}
\bibfield{author}{\bibinfo{person}{Kimberly~E Culley} {and} \bibinfo{person}{Poornima Madhavan}.} \bibinfo{year}{2013}\natexlab{}.
\newblock \showarticletitle{A note of caution regarding anthropomorphism in HCI agents}.
\newblock \bibinfo{journal}{\emph{Computers in Human Behavior}} \bibinfo{volume}{29}, \bibinfo{number}{3} (\bibinfo{year}{2013}), \bibinfo{pages}{577--579}.
\newblock


\bibitem[D'Alfonso et~al\mbox{.}(2020)]%
        {d2020digital}
\bibfield{author}{\bibinfo{person}{Simon D'Alfonso}, \bibinfo{person}{Reeva Lederman}, \bibinfo{person}{Sandra Bucci}, \bibinfo{person}{Katherine Berry}, {et~al\mbox{.}}} \bibinfo{year}{2020}\natexlab{}.
\newblock \showarticletitle{The digital therapeutic alliance and human-computer interaction}.
\newblock \bibinfo{journal}{\emph{JMIR mental health}} \bibinfo{volume}{7}, \bibinfo{number}{12} (\bibinfo{year}{2020}), \bibinfo{pages}{e21895}.
\newblock


\bibitem[De~Choudhury et~al\mbox{.}(2023)]%
        {de2023benefits}
\bibfield{author}{\bibinfo{person}{Munmun De~Choudhury}, \bibinfo{person}{Sachin~R Pendse}, {and} \bibinfo{person}{Neha Kumar}.} \bibinfo{year}{2023}\natexlab{}.
\newblock \showarticletitle{Benefits and Harms of Large Language Models in Digital Mental Health}.
\newblock \bibinfo{journal}{\emph{arXiv preprint arXiv:2311.14693}} (\bibinfo{year}{2023}).
\newblock


\bibitem[De~Waal(2010)]%
        {de2010age}
\bibfield{author}{\bibinfo{person}{Frans De~Waal}.} \bibinfo{year}{2010}\natexlab{}.
\newblock \bibinfo{booktitle}{\emph{The age of empathy: Nature's lessons for a kinder society}}.
\newblock \bibinfo{publisher}{Crown}.
\newblock


\bibitem[Dozois et~al\mbox{.}(2019)]%
        {dozois2019historical}
\bibfield{author}{\bibinfo{person}{David~JA Dozois}, \bibinfo{person}{Keith~S Dobson}, {and} \bibinfo{person}{Katerina Rnic}.} \bibinfo{year}{2019}\natexlab{}.
\newblock \showarticletitle{Historical and philosophical bases of the cognitive-behavioral therapies}.
\newblock \bibinfo{journal}{\emph{Handbook of cognitive-behavioral therapies}} (\bibinfo{year}{2019}), \bibinfo{pages}{3--31}.
\newblock


\bibitem[Durmus et~al\mbox{.}(2023)]%
        {durmus2023towards}
\bibfield{author}{\bibinfo{person}{Esin Durmus}, \bibinfo{person}{Karina Nyugen}, \bibinfo{person}{Thomas~I Liao}, \bibinfo{person}{Nicholas Schiefer}, \bibinfo{person}{Amanda Askell}, \bibinfo{person}{Anton Bakhtin}, \bibinfo{person}{Carol Chen}, \bibinfo{person}{Zac Hatfield-Dodds}, \bibinfo{person}{Danny Hernandez}, \bibinfo{person}{Nicholas Joseph}, {et~al\mbox{.}}} \bibinfo{year}{2023}\natexlab{}.
\newblock \showarticletitle{Towards measuring the representation of subjective global opinions in language models}.
\newblock \bibinfo{journal}{\emph{arXiv preprint arXiv:2306.16388}} (\bibinfo{year}{2023}).
\newblock


\bibitem[Ernala et~al\mbox{.}(2022)]%
        {ernala2022reintegration}
\bibfield{author}{\bibinfo{person}{Sindhu~Kiranmai Ernala}, \bibinfo{person}{Jordyn Seybolt}, \bibinfo{person}{Dong~Whi Yoo}, \bibinfo{person}{Michael~L Birnbaum}, \bibinfo{person}{John~M Kane}, {and} \bibinfo{person}{Munmun De~Choudhury}.} \bibinfo{year}{2022}\natexlab{}.
\newblock \showarticletitle{The Reintegration Journey Following a Psychiatric Hospitalization: Examining the Role of Social Technologies}.
\newblock \bibinfo{journal}{\emph{Proceedings of the ACM on Human-computer Interaction}} \bibinfo{volume}{6}, \bibinfo{number}{CSCW1} (\bibinfo{year}{2022}), \bibinfo{pages}{1--31}.
\newblock


\bibitem[Evans-Lacko et~al\mbox{.}(2018)]%
        {evans2018socio}
\bibfield{author}{\bibinfo{person}{Sarah Evans-Lacko}, \bibinfo{person}{Sergio Aguilar-Gaxiola}, \bibinfo{person}{Ali Al-Hamzawi}, \bibinfo{person}{Jordi Alonso}, \bibinfo{person}{Corina Benjet}, \bibinfo{person}{Ronny Bruffaerts}, \bibinfo{person}{WT Chiu}, \bibinfo{person}{Silvia Florescu}, \bibinfo{person}{Giovanni de Girolamo}, \bibinfo{person}{Oye Gureje}, {et~al\mbox{.}}} \bibinfo{year}{2018}\natexlab{}.
\newblock \showarticletitle{Socio-economic variations in the mental health treatment gap for people with anxiety, mood, and substance use disorders: results from the WHO World Mental Health (WMH) surveys}.
\newblock \bibinfo{journal}{\emph{Psychological medicine}} \bibinfo{volume}{48}, \bibinfo{number}{9} (\bibinfo{year}{2018}), \bibinfo{pages}{1560--1571}.
\newblock


\bibitem[Fogliato et~al\mbox{.}(2022)]%
        {fogliato2022goes}
\bibfield{author}{\bibinfo{person}{Riccardo Fogliato}, \bibinfo{person}{Shreya Chappidi}, \bibinfo{person}{Matthew Lungren}, \bibinfo{person}{Paul Fisher}, \bibinfo{person}{Diane Wilson}, \bibinfo{person}{Michael Fitzke}, \bibinfo{person}{Mark Parkinson}, \bibinfo{person}{Eric Horvitz}, \bibinfo{person}{Kori Inkpen}, {and} \bibinfo{person}{Besmira Nushi}.} \bibinfo{year}{2022}\natexlab{}.
\newblock \showarticletitle{Who goes first? Influences of human-AI workflow on decision making in clinical imaging}. In \bibinfo{booktitle}{\emph{Proceedings of the 2022 ACM Conference on Fairness, Accountability, and Transparency}}. \bibinfo{pages}{1362--1374}.
\newblock


\bibitem[Frank and Frank(1993)]%
        {frank1993persuasion}
\bibfield{author}{\bibinfo{person}{Jerome~D Frank} {and} \bibinfo{person}{Julia~B Frank}.} \bibinfo{year}{1993}\natexlab{}.
\newblock \bibinfo{booktitle}{\emph{Persuasion and healing: A comparative study of psychotherapy}}.
\newblock \bibinfo{publisher}{JHU Press}.
\newblock


\bibitem[Frankl(1985)]%
        {frankl1985man}
\bibfield{author}{\bibinfo{person}{Viktor~E Frankl}.} \bibinfo{year}{1985}\natexlab{}.
\newblock \bibinfo{booktitle}{\emph{Man's search for meaning}}.
\newblock \bibinfo{publisher}{Simon and Schuster}.
\newblock


\bibitem[Freud and Breuer(2004)]%
        {freud2004studies}
\bibfield{author}{\bibinfo{person}{Sigmund Freud} {and} \bibinfo{person}{Joseph Breuer}.} \bibinfo{year}{2004}\natexlab{}.
\newblock \bibinfo{booktitle}{\emph{Studies in hysteria}}.
\newblock \bibinfo{publisher}{Penguin}.
\newblock


\bibitem[Gabriel(2020)]%
        {gabriel2020artificial}
\bibfield{author}{\bibinfo{person}{Iason Gabriel}.} \bibinfo{year}{2020}\natexlab{}.
\newblock \showarticletitle{Artificial intelligence, values, and alignment}.
\newblock \bibinfo{journal}{\emph{Minds and machines}} \bibinfo{volume}{30}, \bibinfo{number}{3} (\bibinfo{year}{2020}), \bibinfo{pages}{411--437}.
\newblock


\bibitem[Gobodo-Madikizela(2004)]%
        {gobodo2004human}
\bibfield{author}{\bibinfo{person}{Pumla Gobodo-Madikizela}.} \bibinfo{year}{2004}\natexlab{}.
\newblock \bibinfo{booktitle}{\emph{A human being died that night: A South African woman confronts the legacy of apartheid}}.
\newblock \bibinfo{publisher}{Houghton Mifflin Harcourt}.
\newblock


\bibitem[Grant and Mandell(2016)]%
        {grant2016boundaries}
\bibfield{author}{\bibinfo{person}{Jill~G Grant} {and} \bibinfo{person}{Deena Mandell}.} \bibinfo{year}{2016}\natexlab{}.
\newblock \showarticletitle{Boundaries and relationships between service users and service providers in community mental health services}.
\newblock \bibinfo{journal}{\emph{Social work in mental health}} \bibinfo{volume}{14}, \bibinfo{number}{6} (\bibinfo{year}{2016}), \bibinfo{pages}{696--713}.
\newblock


\bibitem[Harrer(2023)]%
        {harrer2023attention}
\bibfield{author}{\bibinfo{person}{Stefan Harrer}.} \bibinfo{year}{2023}\natexlab{}.
\newblock \showarticletitle{Attention is not all you need: the complicated case of ethically using large language models in healthcare and medicine}.
\newblock \bibinfo{journal}{\emph{EBioMedicine}}  \bibinfo{volume}{90} (\bibinfo{year}{2023}).
\newblock


\bibitem[Heinlen et~al\mbox{.}(2003)]%
        {heinlen2003nature}
\bibfield{author}{\bibinfo{person}{Kathleen~T Heinlen}, \bibinfo{person}{Elizabeth~Reynolds Welfel}, \bibinfo{person}{Elizabeth~N Richmond}, {and} \bibinfo{person}{Melissa~S O'Donnell}.} \bibinfo{year}{2003}\natexlab{}.
\newblock \showarticletitle{The nature, scope, and ethics of psychologists'e-therapy Web sites: What consumers find when surfing the Web.}
\newblock \bibinfo{journal}{\emph{Psychotherapy: theory, research, practice, training}} \bibinfo{volume}{40}, \bibinfo{number}{1-2} (\bibinfo{year}{2003}), \bibinfo{pages}{112}.
\newblock


\bibitem[Henson et~al\mbox{.}(2019)]%
        {henson2019considering}
\bibfield{author}{\bibinfo{person}{Philip Henson}, \bibinfo{person}{Pamela Peck}, {and} \bibinfo{person}{John Torous}.} \bibinfo{year}{2019}\natexlab{}.
\newblock \showarticletitle{Considering the therapeutic alliance in digital mental health interventions}.
\newblock \bibinfo{journal}{\emph{Harvard review of psychiatry}} \bibinfo{volume}{27}, \bibinfo{number}{4} (\bibinfo{year}{2019}), \bibinfo{pages}{268--273}.
\newblock


\bibitem[Ismail et~al\mbox{.}(2023)]%
        {ismail2023public}
\bibfield{author}{\bibinfo{person}{Azra Ismail}, \bibinfo{person}{Divy Thakkar}, \bibinfo{person}{Neha Madhiwalla}, {and} \bibinfo{person}{Neha Kumar}.} \bibinfo{year}{2023}\natexlab{}.
\newblock \showarticletitle{Public Health Calls for/with AI: An Ethnographic Perspective}.
\newblock \bibinfo{journal}{\emph{Proceedings of the ACM on Human-Computer Interaction}} \bibinfo{volume}{7}, \bibinfo{number}{CSCW2} (\bibinfo{year}{2023}), \bibinfo{pages}{1--26}.
\newblock


\bibitem[Jadhav(2009)]%
        {jadhav2009what}
\bibfield{author}{\bibinfo{person}{Sushrut Jadhav}.} \bibinfo{year}{2009}\natexlab{}.
\newblock \bibinfo{booktitle}{\emph{What is Cultural Validity and Why is it ignored?}}
\newblock \bibinfo{publisher}{AmsterdamAMB}.
\newblock


\bibitem[Jargon(2023)]%
        {jargon2023}
\bibfield{author}{\bibinfo{person}{Julie Jargon}.} \bibinfo{year}{2023}\natexlab{}.
\newblock \showarticletitle{How a Chatbot Went Rogue}.
\newblock \bibinfo{journal}{\emph{Wall Street Journal}} (\bibinfo{year}{2023}).
\newblock
\urldef\tempurl%
\url{https://www.wsj.com/articles/how-a-chatbot-went-rogue-431ff9f9}
\showURL{%
\tempurl}


\bibitem[Jo et~al\mbox{.}(2023)]%
        {jo2023understanding}
\bibfield{author}{\bibinfo{person}{Eunkyung Jo}, \bibinfo{person}{Daniel~A Epstein}, \bibinfo{person}{Hyunhoon Jung}, {and} \bibinfo{person}{Young-Ho Kim}.} \bibinfo{year}{2023}\natexlab{}.
\newblock \showarticletitle{Understanding the benefits and challenges of deploying conversational AI leveraging large language models for public health intervention}. In \bibinfo{booktitle}{\emph{Proceedings of the 2023 CHI Conference on Human Factors in Computing Systems}}. \bibinfo{pages}{1--16}.
\newblock


\bibitem[Jo et~al\mbox{.}(2024)]%
        {jo2024carecall_ltm}
\bibfield{author}{\bibinfo{person}{Eunkyung Jo}, \bibinfo{person}{Yuin Jeong}, \bibinfo{person}{Sohyun Park}, \bibinfo{person}{Daniel~A. Epstein}, {and} \bibinfo{person}{Young-Ho Kim}.} \bibinfo{year}{2024}\natexlab{}.
\newblock \showarticletitle{Understanding the Impact of Long-Term Memory on Self-Disclosure with Large Language Model-Driven Chatbots for Public Health Intervention}. In \bibinfo{booktitle}{\emph{Proceedings of the CHI Conference on Human Factors in Computing Systems}} (Honolulu, HI, USA) \emph{(\bibinfo{series}{CHI '24})}. \bibinfo{publisher}{Association for Computing Machinery}, \bibinfo{address}{New York, NY, USA}, Article \bibinfo{articleno}{440}, \bibinfo{numpages}{21}~pages.
\newblock
\showISBNx{9798400703300}
\urldef\tempurl%
\url{https://doi.org/10.1145/3613904.3642420}
\showDOI{\tempurl}


\bibitem[Kaveladze and Schueller(2023)]%
        {kaveladze2023digital}
\bibfield{author}{\bibinfo{person}{Benjamin Kaveladze} {and} \bibinfo{person}{Stephen~M Schueller}.} \bibinfo{year}{2023}\natexlab{}.
\newblock \showarticletitle{A digital therapeutic alliance in digital mental health}.
\newblock In \bibinfo{booktitle}{\emph{Digital therapeutics for mental health and addiction}}. \bibinfo{publisher}{Elsevier}, \bibinfo{pages}{87--98}.
\newblock


\bibitem[Kaziunas et~al\mbox{.}(2019)]%
        {kaziunas2019precarious}
\bibfield{author}{\bibinfo{person}{Elizabeth Kaziunas}, \bibinfo{person}{Michael~S Klinkman}, {and} \bibinfo{person}{Mark~S Ackerman}.} \bibinfo{year}{2019}\natexlab{}.
\newblock \showarticletitle{Precarious interventions: Designing for ecologies of care}.
\newblock \bibinfo{journal}{\emph{Proceedings of the ACM on Human-Computer Interaction}} \bibinfo{volume}{3}, \bibinfo{number}{CSCW} (\bibinfo{year}{2019}), \bibinfo{pages}{1--27}.
\newblock


\bibitem[Kim et~al\mbox{.}(2024)]%
        {kim2024propile}
\bibfield{author}{\bibinfo{person}{Siwon Kim}, \bibinfo{person}{Sangdoo Yun}, \bibinfo{person}{Hwaran Lee}, \bibinfo{person}{Martin Gubri}, \bibinfo{person}{Sungroh Yoon}, {and} \bibinfo{person}{Seong~Joon Oh}.} \bibinfo{year}{2024}\natexlab{}.
\newblock \showarticletitle{Propile: Probing privacy leakage in large language models}.
\newblock \bibinfo{journal}{\emph{Advances in Neural Information Processing Systems}}  \bibinfo{volume}{36} (\bibinfo{year}{2024}).
\newblock


\bibitem[Kim and Song(2023)]%
        {kim2023communicating}
\bibfield{author}{\bibinfo{person}{Taenyun Kim} {and} \bibinfo{person}{Hayeon Song}.} \bibinfo{year}{2023}\natexlab{}.
\newblock \showarticletitle{Communicating the limitations of AI: the effect of message framing and ownership on trust in artificial intelligence}.
\newblock \bibinfo{journal}{\emph{International Journal of Human--Computer Interaction}} \bibinfo{volume}{39}, \bibinfo{number}{4} (\bibinfo{year}{2023}), \bibinfo{pages}{790--800}.
\newblock


\bibitem[Kocielnik et~al\mbox{.}(2019)]%
        {kocielnik2019harborbot}
\bibfield{author}{\bibinfo{person}{Rafal Kocielnik}, \bibinfo{person}{Elena Agapie}, \bibinfo{person}{Alexander Argyle}, \bibinfo{person}{Dennis~T Hsieh}, \bibinfo{person}{Kabir Yadav}, \bibinfo{person}{Breena Taira}, {and} \bibinfo{person}{Gary Hsieh}.} \bibinfo{year}{2019}\natexlab{}.
\newblock \showarticletitle{HarborBot: a chatbot for social needs screening}. In \bibinfo{booktitle}{\emph{AMIA Annual Symposium Proceedings}}, Vol.~\bibinfo{volume}{2019}. American Medical Informatics Association, \bibinfo{pages}{552}.
\newblock


\bibitem[Laestadius et~al\mbox{.}(2022)]%
        {laestadius2022too}
\bibfield{author}{\bibinfo{person}{Linnea Laestadius}, \bibinfo{person}{Andrea Bishop}, \bibinfo{person}{Michael Gonzalez}, \bibinfo{person}{Diana Illen{\v{c}}{\'\i}k}, {and} \bibinfo{person}{Celeste Campos-Castillo}.} \bibinfo{year}{2022}\natexlab{}.
\newblock \showarticletitle{Too human and not human enough: A grounded theory analysis of mental health harms from emotional dependence on the social chatbot Replika}.
\newblock \bibinfo{journal}{\emph{New Media \& Society}} (\bibinfo{year}{2022}), \bibinfo{pages}{14614448221142007}.
\newblock


\bibitem[Lee and Chew(2023)]%
        {lee2023understanding}
\bibfield{author}{\bibinfo{person}{Min~Hun Lee} {and} \bibinfo{person}{Chong~Jun Chew}.} \bibinfo{year}{2023}\natexlab{}.
\newblock \showarticletitle{Understanding the Effect of Counterfactual Explanations on Trust and Reliance on AI for Human-AI Collaborative Clinical Decision Making}.
\newblock \bibinfo{journal}{\emph{Proceedings of the ACM on Human-Computer Interaction}} \bibinfo{volume}{7}, \bibinfo{number}{CSCW2} (\bibinfo{year}{2023}), \bibinfo{pages}{1--22}.
\newblock


\bibitem[Lee et~al\mbox{.}(2021)]%
        {lee2021human}
\bibfield{author}{\bibinfo{person}{Min~Hun Lee}, \bibinfo{person}{Daniel~P Siewiorek}, \bibinfo{person}{Asim Smailagic}, \bibinfo{person}{Alexandre Bernardino}, {and} \bibinfo{person}{Sergi~Berm{\'u}dez Berm{\'u}dez~i Badia}.} \bibinfo{year}{2021}\natexlab{}.
\newblock \showarticletitle{A human-ai collaborative approach for clinical decision making on rehabilitation assessment}. In \bibinfo{booktitle}{\emph{Proceedings of the 2021 CHI conference on human factors in computing systems}}. \bibinfo{pages}{1--14}.
\newblock


\bibitem[Liu et~al\mbox{.}(2024)]%
        {liu2024confronting}
\bibfield{author}{\bibinfo{person}{Yanchen Liu}, \bibinfo{person}{Srishti Gautam}, \bibinfo{person}{Jiaqi Ma}, {and} \bibinfo{person}{Himabindu Lakkaraju}.} \bibinfo{year}{2024}\natexlab{}.
\newblock \showarticletitle{Confronting LLMs with Traditional ML: Rethinking the Fairness of Large Language Models in Tabular Classifications}. In \bibinfo{booktitle}{\emph{Proceedings of the 2024 Conference of the North American Chapter of the Association for Computational Linguistics: Human Language Technologies (Volume 1: Long Papers)}}. \bibinfo{pages}{3603--3620}.
\newblock


\bibitem[Luborsky et~al\mbox{.}(1975)]%
        {luborsky1975comparative}
\bibfield{author}{\bibinfo{person}{Lester Luborsky}, \bibinfo{person}{Barton Singer}, {and} \bibinfo{person}{Lise Luborsky}.} \bibinfo{year}{1975}\natexlab{}.
\newblock \showarticletitle{Comparative studies of psychotherapies: Is it true that everyone has won and all must have prizes?}
\newblock \bibinfo{journal}{\emph{Archives of general psychiatry}} \bibinfo{volume}{32}, \bibinfo{number}{8} (\bibinfo{year}{1975}), \bibinfo{pages}{995--1008}.
\newblock


\bibitem[Marshall(1996)]%
        {marshall1996sampling}
\bibfield{author}{\bibinfo{person}{Martin~N Marshall}.} \bibinfo{year}{1996}\natexlab{}.
\newblock \showarticletitle{Sampling for qualitative research}.
\newblock \bibinfo{journal}{\emph{Family practice}} \bibinfo{volume}{13}, \bibinfo{number}{6} (\bibinfo{year}{1996}), \bibinfo{pages}{522--526}.
\newblock


\bibitem[McClellan et~al\mbox{.}(2021)]%
        {mcclellan2021impact}
\bibfield{author}{\bibinfo{person}{Chandler McClellan}, \bibinfo{person}{Mir~M Ali}, {and} \bibinfo{person}{Ryan Mutter}.} \bibinfo{year}{2021}\natexlab{}.
\newblock \showarticletitle{Impact of mental health treatment on suicide attempts}.
\newblock \bibinfo{journal}{\emph{The Journal of Behavioral Health Services \& Research}}  \bibinfo{volume}{48} (\bibinfo{year}{2021}), \bibinfo{pages}{4--14}.
\newblock


\bibitem[McGrath et~al\mbox{.}(2023)]%
        {mcgrath2023age}
\bibfield{author}{\bibinfo{person}{John~J McGrath}, \bibinfo{person}{Ali Al-Hamzawi}, \bibinfo{person}{Jordi Alonso}, \bibinfo{person}{Yasmin Altwaijri}, \bibinfo{person}{Laura~H Andrade}, \bibinfo{person}{Evelyn~J Bromet}, \bibinfo{person}{Ronny Bruffaerts}, \bibinfo{person}{Jos{\'e} Miguel~Caldas de Almeida}, \bibinfo{person}{Stephanie Chardoul}, \bibinfo{person}{Wai~Tat Chiu}, {et~al\mbox{.}}} \bibinfo{year}{2023}\natexlab{}.
\newblock \showarticletitle{Age of onset and cumulative risk of mental disorders: a cross-national analysis of population surveys from 29 countries}.
\newblock \bibinfo{journal}{\emph{The Lancet Psychiatry}} \bibinfo{volume}{10}, \bibinfo{number}{9} (\bibinfo{year}{2023}), \bibinfo{pages}{668--681}.
\newblock


\bibitem[Meng et~al\mbox{.}(2023)]%
        {meng2023mediated}
\bibfield{author}{\bibinfo{person}{Jingbo Meng}, \bibinfo{person}{Minjin Rheu}, \bibinfo{person}{Yue Zhang}, \bibinfo{person}{Yue Dai}, {and} \bibinfo{person}{Wei Peng}.} \bibinfo{year}{2023}\natexlab{}.
\newblock \showarticletitle{Mediated Social Support for Distress Reduction: AI Chatbots vs. Human}.
\newblock \bibinfo{journal}{\emph{Proceedings of the ACM on Human-Computer Interaction}} \bibinfo{volume}{7}, \bibinfo{number}{CSCW1} (\bibinfo{year}{2023}), \bibinfo{pages}{1--25}.
\newblock


\bibitem[Merriam and Grenier(2019)]%
        {merriam2019qualitative}
\bibfield{author}{\bibinfo{person}{Sharan~B Merriam} {and} \bibinfo{person}{Robin~S Grenier}.} \bibinfo{year}{2019}\natexlab{}.
\newblock \bibinfo{booktitle}{\emph{Qualitative research in practice: Examples for discussion and analysis}}.
\newblock \bibinfo{publisher}{John Wiley \& Sons}.
\newblock


\bibitem[Milne-Ives et~al\mbox{.}(2020)]%
        {milne2020effectiveness}
\bibfield{author}{\bibinfo{person}{Madison Milne-Ives}, \bibinfo{person}{Caroline de Cock}, \bibinfo{person}{Ernest Lim}, \bibinfo{person}{Melissa~Harper Shehadeh}, \bibinfo{person}{Nick de Pennington}, \bibinfo{person}{Guy Mole}, \bibinfo{person}{Eduardo Normando}, {and} \bibinfo{person}{Edward Meinert}.} \bibinfo{year}{2020}\natexlab{}.
\newblock \showarticletitle{The effectiveness of artificial intelligence conversational agents in health care: systematic review}.
\newblock \bibinfo{journal}{\emph{Journal of medical Internet research}} \bibinfo{volume}{22}, \bibinfo{number}{10} (\bibinfo{year}{2020}), \bibinfo{pages}{e20346}.
\newblock


\bibitem[Miner et~al\mbox{.}(2019)]%
        {miner2019key}
\bibfield{author}{\bibinfo{person}{Adam~S Miner}, \bibinfo{person}{Nigam Shah}, \bibinfo{person}{Kim~D Bullock}, \bibinfo{person}{Bruce~A Arnow}, \bibinfo{person}{Jeremy Bailenson}, {and} \bibinfo{person}{Jeff Hancock}.} \bibinfo{year}{2019}\natexlab{}.
\newblock \showarticletitle{Key considerations for incorporating conversational AI in psychotherapy}.
\newblock \bibinfo{journal}{\emph{Frontiers in psychiatry}}  \bibinfo{volume}{10} (\bibinfo{year}{2019}), \bibinfo{pages}{746}.
\newblock


\bibitem[Moore and Caudill(2019)]%
        {moore2019bot}
\bibfield{author}{\bibinfo{person}{Joshua~R Moore} {and} \bibinfo{person}{Robert Caudill}.} \bibinfo{year}{2019}\natexlab{}.
\newblock \showarticletitle{The bot will see you now: a history and review of interactive computerized mental health programs}.
\newblock \bibinfo{journal}{\emph{Psychiatric Clinics}} \bibinfo{volume}{42}, \bibinfo{number}{4} (\bibinfo{year}{2019}), \bibinfo{pages}{627--634}.
\newblock


\bibitem[Nass et~al\mbox{.}(1994)]%
        {nass1994computers}
\bibfield{author}{\bibinfo{person}{Clifford Nass}, \bibinfo{person}{Jonathan Steuer}, {and} \bibinfo{person}{Ellen~R Tauber}.} \bibinfo{year}{1994}\natexlab{}.
\newblock \showarticletitle{Computers are social actors}. In \bibinfo{booktitle}{\emph{Proceedings of the SIGCHI conference on Human factors in computing systems}}. \bibinfo{pages}{72--78}.
\newblock


\bibitem[Neiman(2021)]%
        {neiman_2021}
\bibfield{author}{\bibinfo{person}{Aaron Neiman}.} \bibinfo{year}{2021}\natexlab{}.
\newblock \bibinfo{title}{{The Typing Cure: Interrogating the Therapeutic Alliance in Australia and Online}}.
\newblock \bibinfo{howpublished}{SPA Biennial 2021: Interrogating Inequalities}.
\newblock


\bibitem[Norcross and Alexander(2005)]%
        {norcross2005primer}
\bibfield{author}{\bibinfo{person}{John~C Norcross} {and} \bibinfo{person}{ERIN~F Alexander}.} \bibinfo{year}{2005}\natexlab{}.
\newblock \showarticletitle{A primer on psychotherapy integration}.
\newblock \bibinfo{journal}{\emph{Handbook of psychotherapy integration}}  \bibinfo{volume}{2} (\bibinfo{year}{2005}), \bibinfo{pages}{3--23}.
\newblock


\bibitem[Okolo et~al\mbox{.}(2021)]%
        {okolo2021cannot}
\bibfield{author}{\bibinfo{person}{Chinasa~T Okolo}, \bibinfo{person}{Srujana Kamath}, \bibinfo{person}{Nicola Dell}, {and} \bibinfo{person}{Aditya Vashistha}.} \bibinfo{year}{2021}\natexlab{}.
\newblock \showarticletitle{“It cannot do all of my work”: community health worker perceptions of AI-enabled mobile health applications in rural India}. In \bibinfo{booktitle}{\emph{Proceedings of the 2021 CHI Conference on Human Factors in Computing Systems}}. \bibinfo{pages}{1--20}.
\newblock


\bibitem[Ouyang et~al\mbox{.}(2022)]%
        {ouyang2022training}
\bibfield{author}{\bibinfo{person}{Long Ouyang}, \bibinfo{person}{Jeffrey Wu}, \bibinfo{person}{Xu Jiang}, \bibinfo{person}{Diogo Almeida}, \bibinfo{person}{Carroll Wainwright}, \bibinfo{person}{Pamela Mishkin}, \bibinfo{person}{Chong Zhang}, \bibinfo{person}{Sandhini Agarwal}, \bibinfo{person}{Katarina Slama}, \bibinfo{person}{Alex Ray}, {et~al\mbox{.}}} \bibinfo{year}{2022}\natexlab{}.
\newblock \showarticletitle{Training language models to follow instructions with human feedback}.
\newblock \bibinfo{journal}{\emph{Advances in neural information processing systems}}  \bibinfo{volume}{35} (\bibinfo{year}{2022}), \bibinfo{pages}{27730--27744}.
\newblock


\bibitem[Pan et~al\mbox{.}(2023)]%
        {pan2023risk}
\bibfield{author}{\bibinfo{person}{Yikang Pan}, \bibinfo{person}{Liangming Pan}, \bibinfo{person}{Wenhu Chen}, \bibinfo{person}{Preslav Nakov}, \bibinfo{person}{Min-Yen Kan}, {and} \bibinfo{person}{William~Yang Wang}.} \bibinfo{year}{2023}\natexlab{}.
\newblock \showarticletitle{On the risk of misinformation pollution with large language models}.
\newblock \bibinfo{journal}{\emph{arXiv preprint arXiv:2305.13661}} (\bibinfo{year}{2023}).
\newblock


\bibitem[Pendse et~al\mbox{.}(2023)]%
        {pendse2023marginalization}
\bibfield{author}{\bibinfo{person}{Sachin~R Pendse}, \bibinfo{person}{Neha Kumar}, {and} \bibinfo{person}{Munmun De~Choudhury}.} \bibinfo{year}{2023}\natexlab{}.
\newblock \showarticletitle{Marginalization and the Construction of Mental Illness Narratives Online: Foregrounding Institutions in Technology-Mediated Care}.
\newblock \bibinfo{journal}{\emph{Proceedings of the ACM on Human-Computer Interaction}} \bibinfo{volume}{7}, \bibinfo{number}{CSCW2} (\bibinfo{year}{2023}), \bibinfo{pages}{1--30}.
\newblock


\bibitem[Pendse et~al\mbox{.}(2022)]%
        {pendse2022treatment}
\bibfield{author}{\bibinfo{person}{Sachin~R Pendse}, \bibinfo{person}{Daniel Nkemelu}, \bibinfo{person}{Nicola~J Bidwell}, \bibinfo{person}{Sushrut Jadhav}, \bibinfo{person}{Soumitra Pathare}, \bibinfo{person}{Munmun De~Choudhury}, {and} \bibinfo{person}{Neha Kumar}.} \bibinfo{year}{2022}\natexlab{}.
\newblock \showarticletitle{From treatment to healing: envisioning a decolonial digital mental health}. In \bibinfo{booktitle}{\emph{Proceedings of the 2022 CHI Conference on Human Factors in Computing Systems}}. \bibinfo{pages}{1--23}.
\newblock


\bibitem[Petrozzino(2021)]%
        {petrozzino2021pays}
\bibfield{author}{\bibinfo{person}{Catherine Petrozzino}.} \bibinfo{year}{2021}\natexlab{}.
\newblock \showarticletitle{Who pays for ethical debt in AI?}
\newblock \bibinfo{journal}{\emph{AI and Ethics}} \bibinfo{volume}{1}, \bibinfo{number}{3} (\bibinfo{year}{2021}), \bibinfo{pages}{205--208}.
\newblock


\bibitem[Reardon(2023)]%
        {reardon2023chatbots}
\bibfield{author}{\bibinfo{person}{Sara Reardon}.} \bibinfo{year}{2023}\natexlab{}.
\newblock \showarticletitle{AI Chatbots Could Help Provide Therapy, but Caution Is Needed}.
\newblock \bibinfo{journal}{\emph{Scientific American}} (\bibinfo{date}{14 June} \bibinfo{year}{2023}).
\newblock
\urldef\tempurl%
\url{https://www.scientificamerican.com/article/ai-chatbots-could-help-provide-therapy-but-caution-is-needed/}
\showURL{%
\tempurl}
\newblock
\shownote{Accessed: 2023-11-03}.


\bibitem[Robertson et~al\mbox{.}(1995)]%
        {robertson1995glocalization}
\bibfield{author}{\bibinfo{person}{Roland Robertson} {et~al\mbox{.}}} \bibinfo{year}{1995}\natexlab{}.
\newblock \showarticletitle{Glocalization: Time-space and homogeneity-heterogeneity}.
\newblock \bibinfo{journal}{\emph{Global modernities}} \bibinfo{volume}{2}, \bibinfo{number}{1} (\bibinfo{year}{1995}), \bibinfo{pages}{25--44}.
\newblock


\bibitem[Robinson et~al\mbox{.}(2023)]%
        {robinson2023chatgpt}
\bibfield{author}{\bibinfo{person}{Nathaniel~R Robinson}, \bibinfo{person}{Perez Ogayo}, \bibinfo{person}{David~R Mortensen}, {and} \bibinfo{person}{Graham Neubig}.} \bibinfo{year}{2023}\natexlab{}.
\newblock \showarticletitle{Chatgpt mt: Competitive for high-(but not low-) resource languages}.
\newblock \bibinfo{journal}{\emph{arXiv preprint arXiv:2309.07423}} (\bibinfo{year}{2023}).
\newblock


\bibitem[Rogers(1957)]%
        {rogers1957necessary}
\bibfield{author}{\bibinfo{person}{Carl~R Rogers}.} \bibinfo{year}{1957}\natexlab{}.
\newblock \showarticletitle{The necessary and sufficient conditions of therapeutic personality change.}
\newblock \bibinfo{journal}{\emph{Journal of consulting psychology}} \bibinfo{volume}{21}, \bibinfo{number}{2} (\bibinfo{year}{1957}), \bibinfo{pages}{95}.
\newblock


\bibitem[Roose(2024)]%
        {roose2024ai}
\bibfield{author}{\bibinfo{person}{Kevin Roose}.} \bibinfo{year}{2024}\natexlab{}.
\newblock \showarticletitle{Can A.I. Be Blamed for a Teen’s Suicide?}
\newblock \bibinfo{journal}{\emph{The New York Times}} (\bibinfo{year}{2024}).
\newblock
\urldef\tempurl%
\url{https://www.nytimes.com/2024/10/23/technology/characterai-lawsuit-teen-suicide.html}
\showURL{%
\tempurl}
\newblock
\shownote{Published Oct. 23, 2024; Updated Oct. 24, 2024}.


\bibitem[Rosenbaum(2002)]%
        {rosenbaum2002typing}
\bibfield{author}{\bibinfo{person}{Joshua Rosenbaum}.} \bibinfo{year}{2002}\natexlab{}.
\newblock \showarticletitle{The Typing Cure}.
\newblock \bibinfo{journal}{\emph{Wall Street Journal}} (\bibinfo{year}{2002}), \bibinfo{pages}{R10}.
\newblock


\bibitem[Rosenzweig(1936)]%
        {rosenzweig1936some}
\bibfield{author}{\bibinfo{person}{Saul Rosenzweig}.} \bibinfo{year}{1936}\natexlab{}.
\newblock \showarticletitle{Some implicit common factors in diverse methods of psychotherapy.}
\newblock \bibinfo{journal}{\emph{American journal of Orthopsychiatry}} \bibinfo{volume}{6}, \bibinfo{number}{3} (\bibinfo{year}{1936}), \bibinfo{pages}{412}.
\newblock


\bibitem[Roudometof(2016)]%
        {roudometof2016glocalization}
\bibfield{author}{\bibinfo{person}{Victor Roudometof}.} \bibinfo{year}{2016}\natexlab{}.
\newblock \bibinfo{booktitle}{\emph{Glocalization: A critical introduction}}.
\newblock \bibinfo{publisher}{Routledge}.
\newblock


\bibitem[Russell and Norvig(2010)]%
        {russell2010artificial}
\bibfield{author}{\bibinfo{person}{Stuart~J Russell} {and} \bibinfo{person}{Peter Norvig}.} \bibinfo{year}{2010}\natexlab{}.
\newblock \bibinfo{booktitle}{\emph{Artificial intelligence a modern approach}}.
\newblock \bibinfo{publisher}{London}.
\newblock


\bibitem[Schick and Sch{\"u}tze(2021)]%
        {schick-schutze-2021-just}
\bibfield{author}{\bibinfo{person}{Timo Schick} {and} \bibinfo{person}{Hinrich Sch{\"u}tze}.} \bibinfo{year}{2021}\natexlab{}.
\newblock \showarticletitle{It{'}s Not Just Size That Matters: Small Language Models Are Also Few-Shot Learners}. In \bibinfo{booktitle}{\emph{Proceedings of the 2021 Conference of the North American Chapter of the Association for Computational Linguistics: Human Language Technologies}}, \bibfield{editor}{\bibinfo{person}{Kristina Toutanova}, \bibinfo{person}{Anna Rumshisky}, \bibinfo{person}{Luke Zettlemoyer}, \bibinfo{person}{Dilek Hakkani-Tur}, \bibinfo{person}{Iz~Beltagy}, \bibinfo{person}{Steven Bethard}, \bibinfo{person}{Ryan Cotterell}, \bibinfo{person}{Tanmoy Chakraborty}, {and} \bibinfo{person}{Yichao Zhou}} (Eds.). \bibinfo{publisher}{Association for Computational Linguistics}, \bibinfo{address}{Online}, \bibinfo{pages}{2339--2352}.
\newblock
\urldef\tempurl%
\url{https://doi.org/10.18653/v1/2021.naacl-main.185}
\showDOI{\tempurl}


\bibitem[Schleider(2023)]%
        {schleider2023little}
\bibfield{author}{\bibinfo{person}{Jessica Schleider}.} \bibinfo{year}{2023}\natexlab{}.
\newblock \bibinfo{booktitle}{\emph{Little Treatments, Big Effects: How to Build Meaningful Moments that Can Transform Your Mental Health}}.
\newblock \bibinfo{publisher}{Robinson}.
\newblock


\bibitem[Schleider et~al\mbox{.}(2022)]%
        {schleider2022randomized}
\bibfield{author}{\bibinfo{person}{Jessica~L Schleider}, \bibinfo{person}{Michael~C Mullarkey}, \bibinfo{person}{Kathryn~R Fox}, \bibinfo{person}{Mallory~L Dobias}, \bibinfo{person}{Akash Shroff}, \bibinfo{person}{Erica~A Hart}, {and} \bibinfo{person}{Chantelle~A Roulston}.} \bibinfo{year}{2022}\natexlab{}.
\newblock \showarticletitle{A randomized trial of online single-session interventions for adolescent depression during COVID-19}.
\newblock \bibinfo{journal}{\emph{Nature Human Behaviour}} \bibinfo{volume}{6}, \bibinfo{number}{2} (\bibinfo{year}{2022}), \bibinfo{pages}{258--268}.
\newblock


\bibitem[Scurto et~al\mbox{.}(2023)]%
        {scurto2023probing}
\bibfield{author}{\bibinfo{person}{Hugo Scurto}, \bibinfo{person}{Thomas Similowski}, \bibinfo{person}{Samuel Bianchini}, {and} \bibinfo{person}{Baptiste Caramiaux}.} \bibinfo{year}{2023}\natexlab{}.
\newblock \showarticletitle{Probing Respiratory Care With Generative Deep Learning}.
\newblock \bibinfo{journal}{\emph{Proceedings of the ACM on Human-Computer Interaction}} \bibinfo{volume}{7}, \bibinfo{number}{CSCW2} (\bibinfo{year}{2023}), \bibinfo{pages}{1--34}.
\newblock


\bibitem[Seo et~al\mbox{.}(2024)]%
        {seo2024chacha}
\bibfield{author}{\bibinfo{person}{Woosuk Seo}, \bibinfo{person}{Chanmo Yang}, {and} \bibinfo{person}{Young-Ho Kim}.} \bibinfo{year}{2024}\natexlab{}.
\newblock \showarticletitle{ChaCha: Leveraging Large Language Models to Prompt Children to Share Their Emotions about Personal Events}. In \bibinfo{booktitle}{\emph{Proceedings of the CHI Conference on Human Factors in Computing Systems}} (Honolulu, HI, USA) \emph{(\bibinfo{series}{CHI '24})}. \bibinfo{publisher}{Association for Computing Machinery}, \bibinfo{address}{New York, NY, USA}, Article \bibinfo{articleno}{903}, \bibinfo{numpages}{20}~pages.
\newblock
\showISBNx{9798400703300}
\urldef\tempurl%
\url{https://doi.org/10.1145/3613904.3642152}
\showDOI{\tempurl}


\bibitem[Sharma et~al\mbox{.}(2024)]%
        {sharma2024facilitating}
\bibfield{author}{\bibinfo{person}{Ashish Sharma}, \bibinfo{person}{Kevin Rushton}, \bibinfo{person}{Inna~Wanyin Lin}, \bibinfo{person}{Theresa Nguyen}, {and} \bibinfo{person}{Tim Althoff}.} \bibinfo{year}{2024}\natexlab{}.
\newblock \showarticletitle{Facilitating self-guided mental health interventions through human-language model interaction: A case study of cognitive restructuring}. In \bibinfo{booktitle}{\emph{Proceedings of the CHI Conference on Human Factors in Computing Systems}}. \bibinfo{pages}{1--29}.
\newblock


\bibitem[Shen et~al\mbox{.}(2024)]%
        {shen2024towards}
\bibfield{author}{\bibinfo{person}{Hua Shen}, \bibinfo{person}{Tiffany Knearem}, \bibinfo{person}{Reshmi Ghosh}, \bibinfo{person}{Kenan Alkiek}, \bibinfo{person}{Kundan Krishna}, \bibinfo{person}{Yachuan Liu}, \bibinfo{person}{Ziqiao Ma}, \bibinfo{person}{Savvas Petridis}, \bibinfo{person}{Yi-Hao Peng}, \bibinfo{person}{Li Qiwei}, {et~al\mbox{.}}} \bibinfo{year}{2024}\natexlab{}.
\newblock \showarticletitle{Towards Bidirectional Human-AI Alignment: A Systematic Review for Clarifications, Framework, and Future Directions}.
\newblock \bibinfo{journal}{\emph{arXiv preprint arXiv:2406.09264}} (\bibinfo{year}{2024}).
\newblock


\bibitem[Shum et~al\mbox{.}(2018)]%
        {shum2018eliza}
\bibfield{author}{\bibinfo{person}{Heung-Yeung Shum}, \bibinfo{person}{Xiao-dong He}, {and} \bibinfo{person}{Di Li}.} \bibinfo{year}{2018}\natexlab{}.
\newblock \showarticletitle{From Eliza to XiaoIce: challenges and opportunities with social chatbots}.
\newblock \bibinfo{journal}{\emph{Frontiers of Information Technology \& Electronic Engineering}}  \bibinfo{volume}{19} (\bibinfo{year}{2018}), \bibinfo{pages}{10--26}.
\newblock


\bibitem[Sitaram et~al\mbox{.}(2023)]%
        {sitaram2023everything}
\bibfield{author}{\bibinfo{person}{Sunayana Sitaram}, \bibinfo{person}{Monojit Choudhury}, \bibinfo{person}{Barun Patra}, \bibinfo{person}{Vishrav Chaudhary}, \bibinfo{person}{Kabir Ahuja}, {and} \bibinfo{person}{Kalika Bali}.} \bibinfo{year}{2023}\natexlab{}.
\newblock \showarticletitle{Everything you need to know about multilingual LLMs: Towards fair, performant and reliable models for languages of the world}. In \bibinfo{booktitle}{\emph{Proceedings of the 61st Annual Meeting of the Association for Computational Linguistics (Volume 6: Tutorial Abstracts)}}. \bibinfo{pages}{21--26}.
\newblock


\bibitem[Slov{\'a}k et~al\mbox{.}(2018)]%
        {slovak2018just}
\bibfield{author}{\bibinfo{person}{Petr Slov{\'a}k}, \bibinfo{person}{Nikki Theofanopoulou}, \bibinfo{person}{Alessia Cecchet}, \bibinfo{person}{Peter Cottrell}, \bibinfo{person}{Ferran Altarriba~Bertran}, \bibinfo{person}{Ella Dagan}, \bibinfo{person}{Julian Childs}, {and} \bibinfo{person}{Katherine Isbister}.} \bibinfo{year}{2018}\natexlab{}.
\newblock \showarticletitle{" I just let him cry... Designing Socio-Technical Interventions in Families to Prevent Mental Health Disorders}.
\newblock \bibinfo{journal}{\emph{Proceedings of the ACM on Human-Computer Interaction}} \bibinfo{volume}{2}, \bibinfo{number}{CSCW} (\bibinfo{year}{2018}), \bibinfo{pages}{1--34}.
\newblock


\bibitem[Smith(2015)]%
        {smith2015qualitative}
\bibfield{author}{\bibinfo{person}{Jonathan~A Smith}.} \bibinfo{year}{2015}\natexlab{}.
\newblock \showarticletitle{Qualitative psychology: A practical guide to research methods}.
\newblock \bibinfo{journal}{\emph{Qualitative psychology}} (\bibinfo{year}{2015}), \bibinfo{pages}{1--312}.
\newblock


\bibitem[Solove(2007)]%
        {solove2007ve}
\bibfield{author}{\bibinfo{person}{Daniel~J Solove}.} \bibinfo{year}{2007}\natexlab{}.
\newblock \showarticletitle{I've got nothing to hide and other misunderstandings of privacy}.
\newblock \bibinfo{journal}{\emph{San Diego L. Rev.}}  \bibinfo{volume}{44} (\bibinfo{year}{2007}), \bibinfo{pages}{745}.
\newblock


\bibitem[Song et~al\mbox{.}(2024)]%
        {song2024exploreself}
\bibfield{author}{\bibinfo{person}{Inhwa Song}, \bibinfo{person}{SoHyun Park}, \bibinfo{person}{Sachin~R Pendse}, \bibinfo{person}{Jessica~Lee Schleider}, \bibinfo{person}{Munmun De~Choudhury}, {and} \bibinfo{person}{Young-Ho Kim}.} \bibinfo{year}{2024}\natexlab{}.
\newblock \showarticletitle{ExploreSelf: Fostering User-driven Exploration and Reflection on Personal Challenges with Adaptive Guidance by Large Language Models}.
\newblock \bibinfo{journal}{\emph{arXiv preprint arXiv:2409.09662}} (\bibinfo{year}{2024}).
\newblock


\bibitem[Staab et~al\mbox{.}(2023)]%
        {staab2023beyond}
\bibfield{author}{\bibinfo{person}{Robin Staab}, \bibinfo{person}{Mark Vero}, \bibinfo{person}{Mislav Balunovi{\'c}}, {and} \bibinfo{person}{Martin Vechev}.} \bibinfo{year}{2023}\natexlab{}.
\newblock \showarticletitle{Beyond memorization: Violating privacy via inference with large language models}.
\newblock \bibinfo{journal}{\emph{arXiv preprint arXiv:2310.07298}} (\bibinfo{year}{2023}).
\newblock


\bibitem[Standal(1954)]%
        {standal1954need}
\bibfield{author}{\bibinfo{person}{Stanley~W Standal}.} \bibinfo{year}{1954}\natexlab{}.
\newblock \emph{\bibinfo{title}{The need for positive regard: A contribution to client-centered theory}}.
\newblock \bibinfo{thesistype}{Ph.\,D. Dissertation}. \bibinfo{school}{University of Chicago, Department of Psychology}.
\newblock


\bibitem[Steindl et~al\mbox{.}(2023)]%
        {steindl2023interplay}
\bibfield{author}{\bibinfo{person}{Stanley~R Steindl}, \bibinfo{person}{Marcela Matos}, {and} \bibinfo{person}{Giancarlo Dimaggio}.} \bibinfo{year}{2023}\natexlab{}.
\newblock \showarticletitle{The interplay between therapeutic relationship and therapeutic technique: The whole is more than the sum of its parts}.
\newblock \bibinfo{journal}{\emph{Journal of Clinical Psychology}} \bibinfo{volume}{79}, \bibinfo{number}{7} (\bibinfo{year}{2023}), \bibinfo{pages}{1686--1692}.
\newblock


\bibitem[Talboy and Fuller(2023)]%
        {talboy2023challenging}
\bibfield{author}{\bibinfo{person}{Alaina~N Talboy} {and} \bibinfo{person}{Elizabeth Fuller}.} \bibinfo{year}{2023}\natexlab{}.
\newblock \showarticletitle{Challenging the appearance of machine intelligence: Cognitive bias in LLMs and Best Practices for Adoption}.
\newblock \bibinfo{journal}{\emph{arXiv preprint arXiv:2304.01358}} (\bibinfo{year}{2023}).
\newblock


\bibitem[Taubenfeld et~al\mbox{.}(2024)]%
        {taubenfeld2024systematic}
\bibfield{author}{\bibinfo{person}{Amir Taubenfeld}, \bibinfo{person}{Yaniv Dover}, \bibinfo{person}{Roi Reichart}, {and} \bibinfo{person}{Ariel Goldstein}.} \bibinfo{year}{2024}\natexlab{}.
\newblock \showarticletitle{Systematic biases in LLM simulations of debates}.
\newblock \bibinfo{journal}{\emph{arXiv preprint arXiv:2402.04049}} (\bibinfo{year}{2024}).
\newblock


\bibitem[Torous and Blease(2024)]%
        {Torous2024}
\bibfield{author}{\bibinfo{person}{John Torous} {and} \bibinfo{person}{Charlotte Blease}.} \bibinfo{year}{2024}\natexlab{}.
\newblock \showarticletitle{Generative artificial intelligence in mental health care: potential benefits and current challenges}.
\newblock \bibinfo{journal}{\emph{World Psychiatry}} \bibinfo{volume}{23}, \bibinfo{number}{1} (\bibinfo{year}{2024}), \bibinfo{pages}{1--2}.
\newblock
\showISSN{1723-8617}
\urldef\tempurl%
\url{https://doi.org/10.1002/wps.21148}
\showDOI{\tempurl}


\bibitem[Tseng et~al\mbox{.}(2023)]%
        {tseng2023understanding}
\bibfield{author}{\bibinfo{person}{Yuan-Chi Tseng}, \bibinfo{person}{Weerachaya Jarupreechachan}, {and} \bibinfo{person}{Tuan-He Lee}.} \bibinfo{year}{2023}\natexlab{}.
\newblock \showarticletitle{Understanding the Benefits and Design of Chatbots to Meet the Healthcare Needs of Migrant Workers}.
\newblock \bibinfo{journal}{\emph{Proceedings of the ACM on Human-Computer Interaction}} \bibinfo{volume}{7}, \bibinfo{number}{CSCW2} (\bibinfo{year}{2023}), \bibinfo{pages}{1--34}.
\newblock


\bibitem[van Heerden et~al\mbox{.}(2023)]%
        {van2023global}
\bibfield{author}{\bibinfo{person}{Alastair~C van Heerden}, \bibinfo{person}{Julia~R Pozuelo}, {and} \bibinfo{person}{Brandon~A Kohrt}.} \bibinfo{year}{2023}\natexlab{}.
\newblock \showarticletitle{Global mental health services and the impact of artificial intelligence--Powered large language models}.
\newblock \bibinfo{journal}{\emph{JAMA psychiatry}} \bibinfo{volume}{80}, \bibinfo{number}{7} (\bibinfo{year}{2023}), \bibinfo{pages}{662--664}.
\newblock


\bibitem[Wainberg et~al\mbox{.}(2017)]%
        {wainberg2017challenges}
\bibfield{author}{\bibinfo{person}{Milton~L Wainberg}, \bibinfo{person}{Pamela Scorza}, \bibinfo{person}{James~M Shultz}, \bibinfo{person}{Liat Helpman}, \bibinfo{person}{Jennifer~J Mootz}, \bibinfo{person}{Karen~A Johnson}, \bibinfo{person}{Yuval Neria}, \bibinfo{person}{Jean-Marie~E Bradford}, \bibinfo{person}{Maria~A Oquendo}, {and} \bibinfo{person}{Melissa~R Arbuckle}.} \bibinfo{year}{2017}\natexlab{}.
\newblock \showarticletitle{Challenges and opportunities in global mental health: a research-to-practice perspective}.
\newblock \bibinfo{journal}{\emph{Current psychiatry reports}}  \bibinfo{volume}{19} (\bibinfo{year}{2017}), \bibinfo{pages}{1--10}.
\newblock


\bibitem[Wampold and Imel(2015)]%
        {wampold2015great}
\bibfield{author}{\bibinfo{person}{Bruce~E Wampold} {and} \bibinfo{person}{Zac~E Imel}.} \bibinfo{year}{2015}\natexlab{}.
\newblock \bibinfo{booktitle}{\emph{The great psychotherapy debate: The evidence for what makes psychotherapy work}}.
\newblock \bibinfo{publisher}{Routledge}.
\newblock


\bibitem[Wang(2017)]%
        {wang2017smartphones}
\bibfield{author}{\bibinfo{person}{Wenhuan Wang}.} \bibinfo{year}{2017}\natexlab{}.
\newblock \showarticletitle{Smartphones as social actors? Social dispositional factors in assessing anthropomorphism}.
\newblock \bibinfo{journal}{\emph{Computers in Human Behavior}}  \bibinfo{volume}{68} (\bibinfo{year}{2017}), \bibinfo{pages}{334--344}.
\newblock


\bibitem[Watson(2012)]%
        {watson2012structural}
\bibfield{author}{\bibinfo{person}{Dennis~P Watson}.} \bibinfo{year}{2012}\natexlab{}.
\newblock \showarticletitle{From structural chaos to a model of consumer support: understanding the roles of structure and agency in mental health recovery for the formerly homeless}.
\newblock \bibinfo{journal}{\emph{Journal of Forensic Psychology Practice}} \bibinfo{volume}{12}, \bibinfo{number}{4} (\bibinfo{year}{2012}), \bibinfo{pages}{325--348}.
\newblock


\bibitem[Weaver(2023)]%
        {Weaver2023}
\bibfield{author}{\bibinfo{person}{Matthew Weaver}.} \bibinfo{year}{2023}\natexlab{}.
\newblock \showarticletitle{{AI chatbot ‘encouraged’ man who planned to kill queen, court told}}.
\newblock \bibinfo{journal}{\emph{The Guardian}} (\bibinfo{date}{6 Jul} \bibinfo{year}{2023}).
\newblock
\urldef\tempurl%
\url{https://www.theguardian.com/uk-news/2023/jul/06/ai-chatbot-encouraged-man-who-planned-to-kill-queen-court-told}
\showURL{%
\tempurl}
\newblock
\shownote{Last modified on 2023-07-06 10:49 EDT}.


\bibitem[Weidinger et~al\mbox{.}(2021)]%
        {weidinger2021ethical}
\bibfield{author}{\bibinfo{person}{Laura Weidinger}, \bibinfo{person}{John Mellor}, \bibinfo{person}{Maribeth Rauh}, \bibinfo{person}{Conor Griffin}, \bibinfo{person}{Jonathan Uesato}, \bibinfo{person}{Po-Sen Huang}, \bibinfo{person}{Myra Cheng}, \bibinfo{person}{Mia Glaese}, \bibinfo{person}{Borja Balle}, \bibinfo{person}{Atoosa Kasirzadeh}, {et~al\mbox{.}}} \bibinfo{year}{2021}\natexlab{}.
\newblock \showarticletitle{Ethical and social risks of harm from language models}.
\newblock \bibinfo{journal}{\emph{arXiv preprint arXiv:2112.04359}} (\bibinfo{year}{2021}).
\newblock


\bibitem[Weizenbaum(1976)]%
        {weizenbaum1976computer}
\bibfield{author}{\bibinfo{person}{Joseph Weizenbaum}.} \bibinfo{year}{1976}\natexlab{}.
\newblock \showarticletitle{Computer power and human reason: From judgment to calculation.}
\newblock  (\bibinfo{year}{1976}).
\newblock


\bibitem[Weizenbaum(1977)]%
        {weizenbaum1977computers}
\bibfield{author}{\bibinfo{person}{Joseph Weizenbaum}.} \bibinfo{year}{1977}\natexlab{}.
\newblock \showarticletitle{Computers as" Therapists"}.
\newblock \bibinfo{journal}{\emph{Science}} \bibinfo{volume}{198}, \bibinfo{number}{4315} (\bibinfo{year}{1977}), \bibinfo{pages}{354--354}.
\newblock


\bibitem[White and Epston(1990)]%
        {white1990narrative}
\bibfield{author}{\bibinfo{person}{Michael White} {and} \bibinfo{person}{David Epston}.} \bibinfo{year}{1990}\natexlab{}.
\newblock \bibinfo{booktitle}{\emph{Narrative means to therapeutic ends}}.
\newblock \bibinfo{publisher}{WW Norton \& Company}.
\newblock


\bibitem[Wiener(1960)]%
        {wiener1960some}
\bibfield{author}{\bibinfo{person}{Norbert Wiener}.} \bibinfo{year}{1960}\natexlab{}.
\newblock \showarticletitle{Some Moral and Technical Consequences of Automation: As machines learn they may develop unforeseen strategies at rates that baffle their programmers.}
\newblock \bibinfo{journal}{\emph{Science}} \bibinfo{volume}{131}, \bibinfo{number}{3410} (\bibinfo{year}{1960}), \bibinfo{pages}{1355--1358}.
\newblock


\bibitem[Winograd(1986)]%
        {winograd1986language}
\bibfield{author}{\bibinfo{person}{Terry Winograd}.} \bibinfo{year}{1986}\natexlab{}.
\newblock \showarticletitle{A language/action perspective on the design of cooperative work}. In \bibinfo{booktitle}{\emph{Proceedings of the 1986 ACM conference on Computer-supported cooperative work}}. \bibinfo{pages}{203--220}.
\newblock


\bibitem[Xiang(2023a)]%
        {xiangtessa2023}
\bibfield{author}{\bibinfo{person}{Chloe Xiang}.} \bibinfo{year}{2023}\natexlab{a}.
\newblock \bibinfo{title}{Eating Disorder Helpline Fires Staff, Transitions to Chatbot After Unionization}.
\newblock
\newblock
\urldef\tempurl%
\url{https://www.vice.com/en/article/n7ezkm/eating-disorder-helpline-fires-staff-transitions-to-chatbot-after-unionization}
\showURL{%
\tempurl}


\bibitem[Xiang(2023b)]%
        {xiangsuicide2023}
\bibfield{author}{\bibinfo{person}{Chloe Xiang}.} \bibinfo{year}{2023}\natexlab{b}.
\newblock \bibinfo{title}{'He Would Still Be Here': Man Dies by Suicide After Talking with AI Chatbot, Widow Says}.
\newblock
\newblock
\urldef\tempurl%
\url{https://www.vice.com/en/article/pkadgm/man-dies-by-suicide-after-talking-with-ai-chatbot-widow-says}
\showURL{%
\tempurl}


\bibitem[Xu et~al\mbox{.}(2021)]%
        {xu-etal-2021-detoxifying}
\bibfield{author}{\bibinfo{person}{Albert Xu}, \bibinfo{person}{Eshaan Pathak}, \bibinfo{person}{Eric Wallace}, \bibinfo{person}{Suchin Gururangan}, \bibinfo{person}{Maarten Sap}, {and} \bibinfo{person}{Dan Klein}.} \bibinfo{year}{2021}\natexlab{}.
\newblock \showarticletitle{Detoxifying Language Models Risks Marginalizing Minority Voices}. In \bibinfo{booktitle}{\emph{Proceedings of the 2021 Conference of the North American Chapter of the Association for Computational Linguistics: Human Language Technologies}}, \bibfield{editor}{\bibinfo{person}{Kristina Toutanova}, \bibinfo{person}{Anna Rumshisky}, \bibinfo{person}{Luke Zettlemoyer}, \bibinfo{person}{Dilek Hakkani-Tur}, \bibinfo{person}{Iz~Beltagy}, \bibinfo{person}{Steven Bethard}, \bibinfo{person}{Ryan Cotterell}, \bibinfo{person}{Tanmoy Chakraborty}, {and} \bibinfo{person}{Yichao Zhou}} (Eds.). \bibinfo{publisher}{Association for Computational Linguistics}, \bibinfo{address}{Online}, \bibinfo{pages}{2390--2397}.
\newblock
\urldef\tempurl%
\url{https://doi.org/10.18653/v1/2021.naacl-main.190}
\showDOI{\tempurl}


\bibitem[Xu et~al\mbox{.}(2023)]%
        {xu2023technology}
\bibfield{author}{\bibinfo{person}{Tian Xu}, \bibinfo{person}{Junnan Yu}, \bibinfo{person}{Dylan~Thomas Doyle}, {and} \bibinfo{person}{Stephen Voida}.} \bibinfo{year}{2023}\natexlab{}.
\newblock \showarticletitle{Technology-Mediated Strategies for Coping with Mental Health Challenges: Insights from People with Bipolar Disorder}.
\newblock \bibinfo{journal}{\emph{Proceedings of the ACM on Human-Computer Interaction}} \bibinfo{volume}{7}, \bibinfo{number}{CSCW2} (\bibinfo{year}{2023}), \bibinfo{pages}{1--31}.
\newblock


\bibitem[Yang et~al\mbox{.}(2024)]%
        {yang2024maqa}
\bibfield{author}{\bibinfo{person}{Yongjin Yang}, \bibinfo{person}{Haneul Yoo}, {and} \bibinfo{person}{Hwaran Lee}.} \bibinfo{year}{2024}\natexlab{}.
\newblock \showarticletitle{MAQA: Evaluating Uncertainty Quantification in LLMs Regarding Data Uncertainty}.
\newblock \bibinfo{journal}{\emph{arXiv preprint arXiv:2408.06816}} (\bibinfo{year}{2024}).
\newblock


\bibitem[Zack et~al\mbox{.}(2024)]%
        {zack2024assessing}
\bibfield{author}{\bibinfo{person}{Travis Zack}, \bibinfo{person}{Eric Lehman}, \bibinfo{person}{Mirac Suzgun}, \bibinfo{person}{Jorge~A Rodriguez}, \bibinfo{person}{Leo~Anthony Celi}, \bibinfo{person}{Judy Gichoya}, \bibinfo{person}{Dan Jurafsky}, \bibinfo{person}{Peter Szolovits}, \bibinfo{person}{David~W Bates}, \bibinfo{person}{Raja-Elie~E Abdulnour}, {et~al\mbox{.}}} \bibinfo{year}{2024}\natexlab{}.
\newblock \showarticletitle{Assessing the potential of GPT-4 to perpetuate racial and gender biases in health care: a model evaluation study}.
\newblock \bibinfo{journal}{\emph{The Lancet Digital Health}} \bibinfo{volume}{6}, \bibinfo{number}{1} (\bibinfo{year}{2024}), \bibinfo{pages}{e12--e22}.
\newblock


\end{thebibliography}
